\documentclass[3p,preprint,12pt]{elsarticle}
\usepackage{graphicx} 
\usepackage{svg}
\usepackage{float} 
\usepackage{subfigure} 
\usepackage{amsmath}
\usepackage{amsfonts,amssymb}
\usepackage{longtable}
\usepackage{hyperref}
\usepackage{doi}
\usepackage[pdf]{graphviz}
\usepackage[utf8]{inputenc}
\usepackage{tabularx}
\usepackage{booktabs}
\usepackage{tabularx}
\usepackage{booktabs}
\usepackage{tikz}
\usepackage{dot2texi}
\usetikzlibrary{graphs, positioning, quotes, shapes.geometric}
\usetikzlibrary{shapes,arrows}
\usetikzlibrary{automata,shapes}
\DeclareUnicodeCharacter{00A0}{~}

\usepackage{xpatch}
\makeatletter
\newcommand*{\addFileDependency}[1]{% argument=file name and extension
  \typeout{(#1)}
  \@addtofilelist{#1}
  \IfFileExists{#1}{}{\typeout{No file #1.}}
}
\makeatother
\xpretocmd{\digraph}{\addFileDependency{#2.dot}}{}{}
\begin{document}
\tikzstyle{startstop} = [rectangle, rounded corners, minimum width = 2cm, minimum height=1cm,text centered, draw = black, fill = red!40]
\tikzstyle{io} = [trapezium, trapezium left angle=70, trapezium right angle=110, minimum width=2cm, minimum height=1cm, text centered, draw=black, fill = blue!40]
\tikzstyle{process} = [rectangle, minimum width=3cm, minimum height=1cm, text centered, draw=black, fill = yellow!50]
\tikzstyle{decision} = [diamond, aspect = 3, text centered, draw=black, fill = green!30]

\tikzstyle{arrow} = [->,>=stealth]

\begin{frontmatter}

    % \title{Elsevier \LaTeX\ template\tnoteref{mytitlenote}}
    \title{ A Novel Perspective Process Simulation Framework Based on Automatic Differentiation}
    % \tnotetext[mytitlenote]{Fully documented templates are available in the elsarticle package on \href{http://www.ctan.org/tex-archive/macros/latex/contrib/elsarticle}{CTAN}.}

    \author[auther1]{Shaoyi Yang}
    
    \author[author2]{Minglei Yang}
    \author[author3]{Wenli Du\corref{cor1}}
    \ead{wldu@ecust.edu.cn}
    % \address[author1]{Shanghai}
    % \address[author2]{Shanghai}
    \cortext[cor1]{Corresponding author}

    % \auther{ Minglei Yang, Wenli Du}
    %address
    % \address{Radarweg 29, Amsterdam}
    % \fntext[myfootnote]{Since 1880.}

    % \author[mymainaddress,mysecondaryaddress]{Elsevier Inc}

    % \author[mysecondaryaddress]{Global Customer Service\corref{mycorrespondingauthor}}
    % \cortext[mycorrespondingauthor]{Corresponding author}

    \address[mymainaddress]{Key Laboratory of Smart Manufacturing in Energy Chemical Process, Ministry of Education, East China University of Science and Technology, Shanghai, 200237, China}
    % \address[mysecondaryaddress]{360 Park Avenue South, New York}

    \begin{abstract}
Thermodynamic and flash equilibrium calculations are the cornerstones of simulation process calculations. The iterative approach, a widely used nonlinear problem-solving technique, relies on derivative calculations throughout the procedure that directly affect the stability and effectiveness of the solution. In this study, we use state-of-the-art automatic differentiation frameworks for thermodynamic calculations to obtain precise derivatives without altering the logic of the algorithm. This contrasts with traditional numerical differentiation algorithms and significantly improves the convergence and computational efficiency of process simulations in contrast to numerical differentiation algorithms. Standard chemical phase equilibrium calculations such as PT, PV, and PH flash are used to evaluate an automated differentiation approach with respect to numerical stability and iteration counts. It is used to evaluate the iteration count. The results of the experiment showed that the automatic differentiation method has a more uniform gradient distribution and requires fewer convergence iterations. The experimental results show that the system shows that the process is more uniform. The gradient distribution and computational convergence curves help to highlight the improvements provided by automatic differentiation. In addition, this method shows greater generalizability and can be used more easily in the calculation of various other chemical simulation modules.

% Thermodynamic and flash equilibrium calculations form the cornerstone of simulation process computations. The iterative approach, a widely-used technique to solve nonlinear problems, relies on derivative calculations throughout the procedure, which directly influence the stability and efficiency of the solution. In this study, we employ a state-of-the-art automatic differentiation framework for thermodynamic calculations to obtain precise derivatives without altering the logic of the algorithm. This contrasts with traditional numerical difference algorithms and significantly enhances the convergence and computational efficiency for process simulation. Standard chemical phase equilibrium calculations, including PT, PV, and PH flash, are used to evaluate the automatic differentiation approach in terms of numerical stability and iteration count. The experimental findings showed that the automatic differentiation method exhibits a more uniform gradient distribution and requires fewer convergence iterations. The gradient distribution and computational convergence curves serve to highlight the improvements offered by the automatic differentiation approach. Moreover, the method exhibits greater generalizability and can be more readily applied to the computation of various other chemical simulation modules.
        
    \end{abstract}


\end{frontmatter}

% \linenumbers
% \ifcase\ShellEscapeStatus
%     No shell escape\or
%     Unrestricted shell escape\or
%     Restricted shell escape%
%   \fi

\section*{Notation}

\begin{longtable}[htbp!]{l|lll}
\toprule
Symbol&Meaning&\\
\midrule
$p$ & Pressure&  \\
$R$ & Gas constant & \\
$T$ & Temperature & \\
$V$ & Volume &  \\
$b$ & Molecular volume & \\
$a(T)$ &  Intermolecular attraction& \\
$a_c$ &  Critical intermolecular attraction& \\
$T_r$ & Contrast temperature& \\
$T_c$& Critical temperature& \\
$p_c$& Critical pressure \\
$\alpha(T_r)$& Temperature-dependent dimensionless numbers & \\
$\kappa $ & Slope of $\alpha$ & \\
$\omega $& Acentric factor & \\
$a_m$& Parameters of the mixture a & \\
$y_i$& Molar fraction of components & \\
$y_j$& Molar fraction of components & \\
$a_i$& Intermolecular attraction of components & \\
$a_j$& Intermolecular attraction of components& \\
$k_{ij}$& Binary interaction coefficient & \\
$b_m$ &  Parameters of the mixture b& \\
$b_i$ &  Molecular volume of components& \\
$A$& Intermediate variable & \\
$B$& Intermediate variable & \\
$z$& Compressibility factor & \\
$z_{liq}$& Liquid phase compressibility factor & \\
$z_{vap}$& Vapor phase Compressibility factor & \\
$\Phi_{liq,i}$& Liquid phase fugacity coefficient  & \\
$\Phi_{vap,i}$& Vapor phase fugacity coefficient & \\
$K_i$& Phase equilibrium constant of components&  \\
$a_{ci}$ & Critical intermolecular attraction &  \\
$T_{ri}$ & Contrast temperature of components & \\
$F$ & Total feed flow rate f&  \\
$f_i$ & Feed flow rate rof components &  \\
$H_f$ & Feed enthalpy & \\
$Q$ & Heat duty & \\
$H_L$ &  Liquid enthalpy \\
$H_V$ &  Vapor enthalpy& \\
$z_mix$ &  Mixture phase composition & \\
$x_l$ &  Liquid phase composition& \\
$y_v$ &  Vapor phase composition& \\
$H_total$ & Total enthalpy& \\
$H_{mix}$ &  Mixture phase enthalpy & \\
$H_{out}$ &  Total out enthalpy & \\
$H_{outL}$ &  Liquid phase out enthalpy& \\
$H_{outV}$ &  Vapor phase out enthalpy& \\
$V_{out}$ &  Vapor phase flow rate& \\
$H_{mixL}$ &  Liquid phase enthalpy of input & \\
$H_{mixV}$ &  Vapor phase enthalpy of input & \\
$H_{error}$ &  Iterative enthalpy difference& \\
\bottomrule
\end{longtable}

\section{Introduction}
Process simulation is a powerful tool for analyzing and predicting the behavior of complex systems, such as chemical processes, power plants, and other manufacturing systems. The calculation of process simulations is highly non-linear, with highly constrained and tightly coupled variables. Currently, the main methods are Sequential Modular (SM) and Equation Oriented (EO), which rely heavily on iterative computation.  To ensure the computational efficiency and precision of the algorithm, both of these procedures need derivatives.  As the size and dynamic nature of the issue increase, the precision of the derivative computation and resource overhead of the calculation procedure become increasingly crucial.\par
In chemical process calculations, four principal methods are utilized for calculating derivatives. Manually calculating the formula and directly programming the computation (hand-coded derivatives) are the most straightforward methods. Chemical calculations that do not include nested loops, such as diffusion processes, reaction rate calculations, and heat transfer calculations, can be expressed with explicit derivative formulae in some traditional textbooks\cite{seader2016separation,greenkorn2018momentum}. Researchers often spend effort on the manual derivation of analytical derivatives and then use standard optimization procedures, such as Newton,L-BFGS \cite{L-BFGS} to solve equations. However, this approach has exact derivatives without truncation error and becomes extremely complicated and prone to errors when implicit computations and large-scale equations are involved.\par
Another common method is numerical approximation of derivatives by finite difference, which is simple to develop and utilize existing subroutines for the computation of function values. Almost all implicit iterative computations in process calculations, including phase equilibrium calculations, process loop iterations, and design prescription calculations, can employ the finite-difference approach to calculate the derivatives. However, the greatest disadvantage of the finite difference method is the imprecision due to the rounding and truncation error\cite{jerrell1997automatic}, particularly in chemical engineering calculations. The range of values for different variables varies substantially, and such difficulties will be examined in depth in the following sections.\par
Computational complexity is another shortcoming of the finite difference method. In general, finite-difference approximations of derivatives require at least calculating the original function twice at each point, which is included $2mn$ times evaluations when computing the Jacobian matrix of a function $f:\mathbb{R}^n \to \mathbb{R}^m$. This problem will be more pronounced when higher-order derivatives are required.\par
Symbolic differentiation addresses the shortcomings of numerical differentiation and manual differentiation by using symbolic computation and algebraic algorithms (such as polynomial algorithms and simplification algorithms) to obtain the exact solution for expressions \cite{symblic}. Numerous current toolkits, such as MACSYMA\cite{pavelle1985macsyma}, Maple\cite{char2013maple}, Mathematica, can make symbolic differentiation straightforward, which reduces the complexity of problem solving and makes computation more clear and easy. Tolsma\cite{tolsma1998computational} applied symbolic differentiation in chemical calculations and compared to the other methods and analyzed the advantages of symbolic calculation from the perspective of accuracy and calculation speed.
SyPSE\cite{zhang2021sypse} presented a structure for turning chemical models into symbolic calculations by using computer algebraic concepts for direct solution. However, the exact solution is too
45 complex to store and compute for some complex or nested expressions, which is known as the "expression swell"\cite{expression-swell}. This issue severely restricts the general application of symbolic differentiation.\par

Automatic differentiation (AD)\cite{corliss2002automatic}is a powerful technique that overcomes the weaknesses of the aforementioned methods. AD uses computational graphs to describe expressions for the calculation process which is able to compute the derivative of each equation precisely, eliminating the influence of truncation error on the findings and minimizing round-off errors. AD allows a computer to automatically calculate the derivative of the input of a program with respect to the output, without additional knowledge or human intervention. Unlike the effort involved in using lexical analysis and semantic constraints in symbolic differentiation, AD expands expressions into a computational graph, which can be applied to regular code with minimal modification and can handle branching, loops, and recursion \cite{baydin2018automatic}. \par
AD has been applied to numerical simulations in industry and academia. In physical modeling exact derivatives obtained by AD make calculations more efficient; which has been efficiently applied to calculate the density functionals \cite{eks} and to solve the multigrid cell-vertex Euler flow \cite{AD_in_CFD}. Moreover, Walther \cite{walther2007automatic_in_control} demonstrated that AD has significant advantages in terms of convergence rate and computational stability in optimal control problems. Carmichael\cite{carmichael1997sensitivity} used AD in the sensitivity analysis for atmospheric chemistry models. Haase\cite{haase2002optimal} integrated AD and sequential quadratic programming (SQP) in optimum sizing issues in industrial structural mechanics. Almost everywhere in scientific computing, automatic differentiation can lead to significant improvements, which also shows great potential to solve process simulation problems. \par
In the past decade, all deep learning\cite{lecun2015deep} computations have relied on automatic differentiation and computation graphs. The backpropagation\cite{lecun1988theoretical} algorithm is the most efficient way to train neural networks, AD fits the need for the backpropagation computational model, which has been one of the most studied and widely used training algorithms since its proposal by Rumelhart\cite{rumelhart1986learning_bp}. In the machine learning community, modern code frameworks such as TensorFlow \cite{abadi2016tensorflow}, Autograd \cite{maclaurin2016modeling} and PyTorch \cite{paszke2019pytorch} have brought general purpose AD to the mainstream.\par
Although AD has been widely used in machine learning and other domains, it is not widely used in chemical process simulation. Chan and Prince\cite{chan1986application} used the chain rule in flowsheet calculations for the first time and showed that the result was more accurate than the perturbation-based differencing approach.
Chen and Stadtherr\cite{chen1985simultaneous} also proposed that the use of analytical derivatives was the most suitable technique to provide sensitivity information in their early sequential modular simulator, although they noted that (at the time) the computational cost was often prohibitive. Wolbert\cite{wolbert1994flowsheet} demonstrated that inexact derivatives can lead to failure to compute correct Newton steps in process optimization problems, and moreover described an implementation of analytical derivative evaluation using the chain rule and implicit function sensitivity analysis in a sequential modular process simulator to prove its efficacy. For dynamic simulation Castro\cite{castro2000automatic} applied AD in the DAE system and proposed that numerical differentiation is not suitable for the calculation of derivatives with higher order. In the case of the equation-oriented method, due to the use of iterative solving algorithms, the solution of large-scale equations relies more heavily on derivatives, Li\cite{li2004module} proposed a module-oriented automatic differentiation approach by exploiting the sparsity of the model. Dowling\cite{dowling2015framework} employed the automatic differentiation to obtain multiorder derivatives of large-scale equations. Although some problems have been perfectly addressed in the past, research into thermodynamic calculations through automatic differentiation is still lacking.\par

In this paper, a state-of-the-art automatic differentiation framework is applied to thermodynamic calculations to obtain exact derivatives without modifying the logic of the algorithm, in contrast to traditional numerical difference algorithms, to improve the convergence and computational efficiency of process simulation calculations.
This paper is organized as follows: Section \ref{section2} describes the principle of automatic differentiation and the latest implementation. Section \ref{section3} provides an automatic differentiation and compute graph implementation for the classical Soave-Redlich-Kwong classical thermodynamic property approach and then explains the benefit of the automatic differentiation technique in PT, PV and PH flash, as well as the convergence feature for nested computations. The next section, Section \ref{section4}, discusses potential optimization directions.
\section{Methods and implementations}\label{section2}

\subsection{Irreconcilable contradictions in numerical differentiation}
\begin{figure}[H] 
    \centering 
    \includegraphics[width=0.5\textwidth]{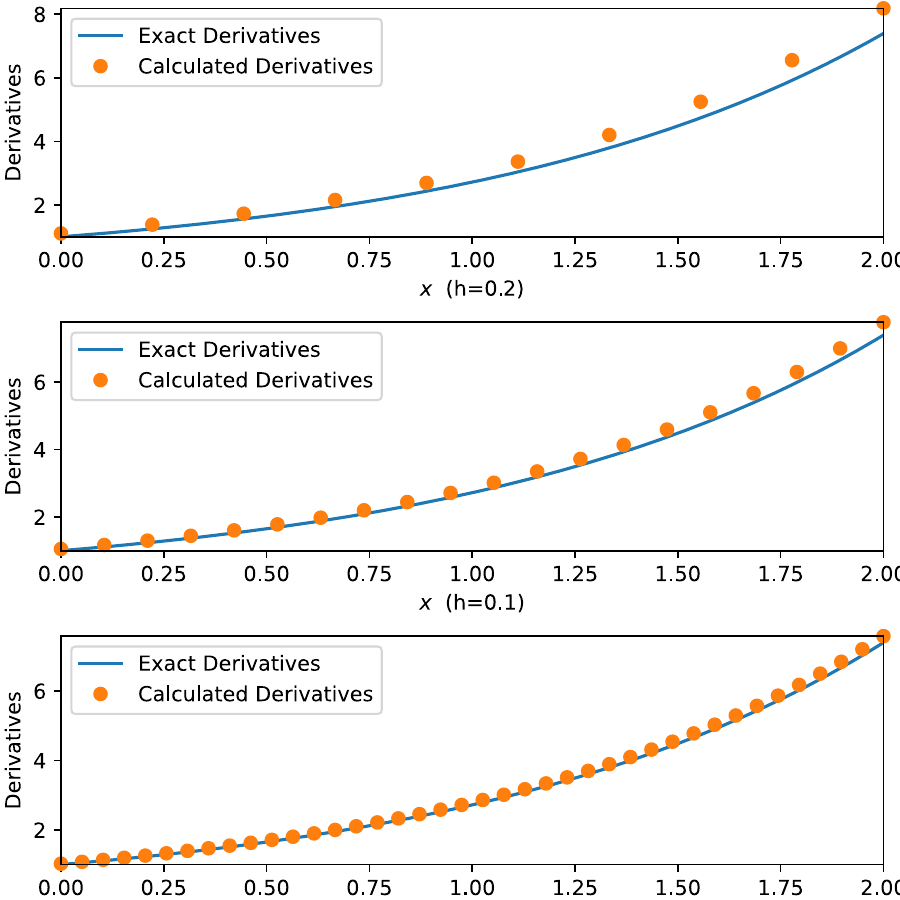} 
    \caption{Numerical differentiation truncation errors} 
    \label{Fig.main2} 
\end{figure}

Numerical algorithms always have round-off and truncation errors when calculating floating point numbers in computers. The traditional method of obtaining approximate numerical derivatives of a function $f$ is through the use of finite differences like equation(\ref{e1}) . As $h$ decreases, the truncation error will gradually decrease, as shown in figure(\ref{Fig.main2}) illustrates the numerical error.
\begin{gather}
\label{e1}
D_+f(x) = \frac{f(x+h)-f(x)}{h} \quad or\quad D_{\pm}f(x) = \frac{f(x+h)-f(x-h)}{2h}
\end{gather}
Although a smaller step size reduces the truncation error, the effect of round-off error becomes more significant as the step size decreases. Although the finite difference technique can handle the calculation of the derivative of the chemical simulation process, there is an issue with irreconcilable errors in the finite difference method calculation. In this sense, the automatic differentiation is proposed to solve the  calculation accuracy problem.
\subsection{Principle of automatic differentiation}
All numerical computations are ultimately compositions of a finite set of elementary operators which the derivatives are known \cite{griewank2008evaluating}. AD utilizes the idea that every function or expression executes a succession of elementary arithmetic operations (such as addition, subtraction, multiplication, and division) and elementary functions, regardless of their complexity (such as exp, log, sin, cos, etc.). By applying the chain rule to these procedures, it is possible to precisely compute derivatives of arbitrary order. This indicates that algorithmic differentiation is not susceptible to truncation mistakes.\par

Forward mode and reverse mode are the fundamental model for automatic differentiation. 
This section demonstrates how the two modes differentiate the output variable $y$ with regard to $x$, as illustrated by the $y$ equation and its derivatives:
\begin{gather}
    y = u_n(u_{n-1}(u_{n-2}...(u_1(x))))\\
    \frac{d y}{d x}=\frac{d y}{d v_n} \frac{d v_n}{d x}=\frac{d y}{d v_n} \frac{d v_n}{d v_{n-1}} \frac{d v_{n-1}}{d x}=\frac{d y}{d v_n} \frac{d v_n}{d v_{n-1}} \ldots \frac{d v_2}{d v_1} \frac{d v_1}{d x}
\end{gather}\par
% Forward mode means calculating the value of the derivative of every one of the variables with respect to the independent variable $x$, in actual calcation it will firstly calcualte the $\frac{d v_1}{dx}$ then $\frac{d v_2}{dv_1}$. 

Forward mode entails calculating the value of the derivative of each variable with respect to the independent variable $x$; in actual calculation, the $\frac{d v_1}{dx}$ and $\frac{d v_2}{dv_1}$ are calculated first.
% The order of calculation of the derivative is the same as the order of its value with the function, and the value of the function required for the derivative can be found together with the corresponding differentiation. The forward mode calculates the differential while computing the forward propagation of the graph, therefore one forward propagation of the forward mode is sufficient to calculate the output and derivative values. Forward mode can get the more exact derivates, but the biggest drawback of the mode is the high time complexity, when input variable has $n$ dimensions, calculating the derivative requires traversing the computational graph $n$ times.\\
In forward-mode automatic differentiation, the computation of the function value and its derivative can be carried out simultaneously during a single forward pass through the computational graph. This is because the order in which derivatives are calculated aligns with the order of computing the function's value. Thus, a single traversal of the graph is sufficient for obtaining both the output of the function and its corresponding derivatives. The major drawback of forward mode is its high temporal complexity; when the input variable has $n$ dimensions, computing the derivative needs $n$ iterations of traversing the computational network.\par
Different from forward mode, Reverse mode  will first calculate the expression$\frac{dy}{dv_n}$, then the $\frac{dv_n}{dv_{n-1}}$. This is a space-for-time calculation mode, the middle result of $u 1(x), u 2(u 1(x))...y$ should be stored in memory. In the Reverse mode, all partial derivatives are computed using a single backward transfer, and intermediate partial derivatives are only computed once. In the case of many parameters, the effort of repetitive calculations is minimized, and the temporal complexity of backward auto-differentiation is lowered. When the output dimension is smaller than the input dimension, the reverse mode requires fewer multiplications than the forward mode. When the output dimension is larger than the input, forward mode differentiation is appropriate; when the output dimension is considerably less than the input, reverse mode differentiation is appropriate.
\subsection{Implementation for automatic differentiation}
There are three primary approaches to automatic differentiation. The simplest among them employs a basic expression method, wherein elementary mathematical operations and their corresponding differential expressions serve as library functions. During code execution, each basic expression and its combinations are recorded. The derivatives are then computed using the chain rule during the code generation phase. DiffSharp\cite{baydin2015diffsharp} is a representative example that utilizes this method based on a library of basic expressions.The second approach leverages operator overloading (OO), taking advantage of language polymorphism. This method modifies the behavior of basic operators, recording each elementary expression and its combinations during code execution. The derivatives are then directly generated according to the chain rule. This approach offers easier implementation and allows for more lightweight development compared to the basic expression method. PyTorch serves as a quintessential example of automatic differentiation through operator overloading, utilizing dynamic computational graphs.\par
Previous frameworks for automatic differentiation were adept at handling mathematical functions but fell short when dealing with complex mathematical logic and exhibited suboptimal execution performance. The most versatile method for implementing automatic differentiation currently available is based on source code transformations. This technique relies on functional programming and performs automatic differentiation by analyzing the code's abstract syntax tree (AST). It supports complex control flow scenarios, thereby enhancing the versatility of automatic differentiation. Given the intricate logic and conditional judgments involved in chemical process simulation calculations, the source-code-transformation-based method is particularly well-suited for these computational requirements.
In this study, we employ JAX\cite{frostig2018compiling}, a cutting-edge automatic differentiation framework that relies on source code transformations. JAX features a domain-specific tracing JIT compiler capable of generating high-performance accelerator code from pure Python. We leverage JAX to seamlessly integrate automatic differentiation into our process simulation calculations.
% \subsection{Computation Graph optimization}
% The computational graph is a tool used to represent the logic and state of a computation. The computational framework represents the computation and backward gradient computation in the form of a computational graph in the backend. A computational graph consists of a basic data structure: a tensor and a basic unit of operation: an operator. In a computational graph, nodes are usually used to represent operators, and directed line segments between nodes represent tensor states, and also describe the dependencies between computations. The data flow will update the tensor state in the graph by performing forward and backward gradient calculations with operators according to the flow direction in the graph.
% \section{Differentiable thermodynamic model}\label{section3}
\section{Result and Discussion}\label{section3}
% Thermodynamic model is the basis of chemical process simulation, and choosing an accurate and efficient thermodynamic property package is the key to accurate modeling. Almost all process simulation needs accurate property to calculate phase Equilibrium. Differentiable thermodynamic model means using computational graph represents the mixing rules and the calculation of thermodynamic property, when in flash calculation or other calculation needs thermodynamic property derivatives it will use automatic differentiation obtain accurate derivatives without truncation errors, and in flowsheet optimization differentiable thermodynamic model can give derivatives for flowsheet solving.
The thermodynamic model is the foundation of chemical process simulation, and selecting a precise and efficient thermodynamic property package is the key to correct modeling. Almost any process simulation needs correct property to compute phase Equilibrium. Differentiable thermodynamic model refers to the use of computational graphs to express mixing rules and calculate thermodynamic parameters. When flash calculations or other computations need thermodynamic property derivatives, automated differentiation is employed to generate exact derivatives that are free of truncation errors. Using operational conversion, the differentiable thermodynamic model constructs an auto-differentiation-enabled property package. This model gives a computational graph of the computational process, optimizes the computational flow of the property calculation, adapts it to massively parallel computing, and substantially enhances computational efficiency. The greatest benefit of this model is that it gives a continuously differentiable approach for estimating thermodynamic characteristics, which enables the efficient acquisition of precise derivatives. The precise derivatives considerably increase the efficiency of model iteration and prevent failures in process convergence caused by numerical inaccuracies and discontinuities in numerical derivatives.\par

% The differentiable thermodynamic model uses operational conversion constructing an auto-differentiation-enabled property package. This model provides a computational graph of the computational process, optimizes the computational flow of the property calculation, adapts it to massively parallel computing, and significantly improves computational efficiency. The most important advantage of this model is that provides a continuously differentiable method for calculating thermodynamic properties, which facilitate the efficient acquisition of accurate derivatives. The accurate derivatives significantly improve model iteration efficiency and prevents process convergence failures due to numerical errors and discontinuities in numerical derivatives.
\subsection{Differentiable Thermodynamic Models}
In this paper we use the SRK equation of state as a property package for the key components of the ethylene production process (methane, ethylene, ethane, propane), another reason for choosing SRK is that it involves the selection of the roots of cubic equations, an automatic differentiation process involving logical judgments that has been difficult to implement in past tools.
% Using operational conversion, the differentiable thermodynamic model constructs an auto-differentiation-enabled property package. This model gives a computational graph of the computational process, optimizes the computational flow of the property calculation, adapts it to massively parallel computing, and substantially enhances computational efficiency. The greatest benefit of this model is that it gives a continuously differentiable approach for estimating thermodynamic characteristics, which enables the efficient acquisition of precise derivatives. The precise derivatives considerably increase the efficiency of model iteration and prevent failures in process convergence caused by numerical inaccuracies and discontinuities in numerical derivatives.
% \subsection{Base of SRK Model}

The form of the SRK equation is
% \begin{equation}
%     p = \frac{RT}{V-b_m}-\frac{a(T)}{V(V+b_m)}
%     a(T) = a_c *\alpha(T_r)=0.42728*\frac{R^2T_c^{2}}{p_c}* \alpha(T_r)}
% \end{equation}
\begin{gather}
    % \begin{align}
    p                = \frac{RT}{V-b}-\frac{a(T)}{V(V+b)}  \label{SRKfunc}          \\
    a(T)             = a_c *\alpha(T_r)=0.42728*\frac{R^2T_c^{2}}{p_c}* \alpha(T_r) \\
    % a_c = 0.42748*\frac{R^2T_c^2}{p_c}\\
    b_c                = 0.08664*\frac{RT_c}{p_c}                                      \\
    \alpha(T_r)^{0.5}  = 1+\kappa*(1-T_r^{0.5})\\
    \kappa = 0.48+1.574*\omega-0.176*\omega^2
    % y2=x2+z2 \label{YY}        \\
    % y3=x3+z3 \tag{\ref{YY}{a}} \\
    % y3=x3+z3 \notag            \\
    % y4=x4+z4 \tag{\ref{YY}{b}} \label{YYb}
    % \end{align}
\end{gather}\par
When function (\ref{SRKfunc}) applied to the mixture, the equation of state parameters $a_m$ and $b_m$ can be derived from the pure component parameter $a_i$ and $b_i$, the mixing rule for $a_m, b_m$ can be expressed as($y_i$ denotes component fraction):\par

\begin{gather}
    a_m = \sum_{i}\sum_{j}y_iy_j(a_i*a_j)^{0.5}(1-k_{ij})\\
    b_m = \sum_{i}y_ib_i\\
    A = \frac{a_mp}{R^2T^2}\\
    B = \frac{b_m}{RT}
\end{gather}\par
Using the compressibility factor to represent the SRK equation, the pure component fugacity coefficient derived from it.
\begin{gather}
    z^3-z^2+z(A-B-B^2)-AB=0\label{z}\\
    z_{liq} = min(z)\\
    z_{vap} = max(z)\\
    \phi_{liq,i} = exp((z_{liq}-1)-ln(z_{liq}-B)-\frac{A}{B}ln(1+\frac{B}{z_{liq}}))\label{phi_liq}\\
    \phi_{vap,i} = exp((z_{vap}-1)-ln(z_{vap}-B)-\frac{A}{B}ln(1+\frac{B}{z_{vap}}))\label{phi_vap}\\
    \frac{h-h^0}{RT} = (z-1)-ln(1+\frac{B}{z})(\frac{A}{B}+\frac{\sqrt{a}}{RTb}\sum_{i}x_i\kappa_i\sqrt{a_{ci}T_{ri}})\\
    K_{i} = \frac{\phi_{liq,i}}{\phi_{vap,i}}
\end{gather}\par

In calculating the fugacity coefficient, the first step is to solve the cubic equation, and in this part we use the Shingin formula\cite{fan1989new} to obtain a more explicit expression by conditional judgment. In previous research, Kamath\cite{kamath2010equation} employed first-order and second-order derivative data to constrain the cubic equation, facilitating the acquisition of the essential gas-liquid phase compression factor. This approach, however, may prove to be less effective when dealing with more intricate equation of state (EOS) functions that do not exhibit a clear positive or negative correlation between variables. Consequently, determining the appropriate roots for such functions could become increasingly challenging.

In our study, we addressed this issue by formulating distinct equations for the calculation of the two roots separately. We then implemented automatic differentiation to expand these equations, effectively circumventing the need to ascertain the roots. This alternative method not only streamlines the process but also ensures the accurate determination of roots for complex EOS functions.

The phase equilibrium constant K is the core of chemical process calculation, the ideal simplified method uses the temperature pressure to calculate the equilibrium constant, the rigorous calculation method uses the vapor phase and liquid phase fugacity. In the next flash calculation, the conventional algorithm adjusts the phase equilibrium constant K by iterative temperature. The derivative of the phase equilibrium constant K with respect to temperature will improve the efficiency of the iterative algorithm. Due to the multi-layer nested computational structure, numerical errors can be significant when using the finite difference method to obtain numerical derivatives. Figure(\ref{figureK}) illustrates the calculation logic of the phase equilibrium constant using the form of nodes.

\begin{figure}[H]
    \centering
    \begin{dot2tex}[fdp,mathmode,graphstyle={scale=0.5,transform shape}]
        digraph G {
        node [shape="record"];
        edge [lblstyle="auto"];
        Lfrac [label="L_{frac}", style="state, initial"]
        Vfrac [label="V_{frac}", style="state, initial"]
        T [style="state, initial"]
        P [style="state, initial"]
        Lfug [label="L_{Fug}"];
        Vfug [label="V_{Fug}"];
        Lfrac->Mixrule
        Vfrac->Mixrule
        T->Mixrule->Lfug->K
        P->Mixrule->Vfug->K;
        K [style="state,accepting"];
        }
    \end{dot2tex}
    \caption{Computational Graph of K}
    \label{figureK}
\end{figure}
\subsection{Exact calculation of the derivative of K}
Differentiable thermodynamic model using automatic differentiation to calculate the exact derivatives of K with respect to other parameters. In general flash calculations, the effects of temperature, pressure and component fraction on the phase equilibrium constants are usually of interest.
The temperature acts on the calculation of the phase equilibrium constant K by influencing the compression factor and parameters A, B in the equation of the fugacity coefficient, figure(\ref{figureK-T}) illustrates its computational dependence.
\begin{figure}[H]
    \centering
    \begin{dot2tex}[dot,options=-tmath,graphstyle={scale=0.6,transform shape}]
        digraph G {
        node [shape="record"];
        // edge [lblstyle="auto",topath="bend left"];
        edge [lblstyle="auto"];
        dKdphiv [label="\\frac{\\partial K}{\\partial \\phi_{vap}}"]
        dphivdzv [label="\\frac{\\partial \\phi_{vap}}{\\partial z_v}"]
        dphivdAv [label="\\frac{\\partial \\phi_{vap}}{\\partial A_v}"]
        dphivdBv [label="\\frac{\\partial \\phi_{vap}}{\\partial B_v}"]

        dKdphil [label="\\frac{\\partial K}{\\partial \\phi_{liq}}"]
        dphildzl [label="\\frac{\\partial \\phi_{liq}}{\\partial z_l}"]
        dphildAl [label="\\frac{\\partial \\phi_{liq}}{\\partial A_l}"]
        dphildBl [label="\\frac{\\partial \\phi_{liq}}{\\partial B_l}"]

        dzldAl [label="\\frac{\\partial z_l}{\\partial A_l}"]
        dzldBl [label="\\frac{\\partial z_l}{\\partial B_l}"]
        dAldT [label="\\frac{\\partial A_l}{\\partial T}"]
        dBldT [label="\\frac{\\partial B_l}{\\partial T}"]
        dKdT [label="\\frac{\\partial K}{\\partial T}",style="state, initial"]

        dzvdAv [label="\\frac{\\partial z_v}{\\partial A_v}"]
        dzvdBv [label="\\frac{\\partial z_v}{\\partial B_v}"]
        dAvdT [label="\\frac{\\partial A_{l}}{\\partial T}"]
        dBvdT [label="\\frac{\\partial B_{l}}{\\partial T}"]

        dKdT->dKdphil[topath="bend left"]
        dKdphil->dphildzl
        dphildzl->dzldAl
        dzldAl->dAldT

        dphildzl->dzldBl
        dzldBl->dBldT

        dKdphil->dphildAl
        dphildAl->dAldT
        dKdphil->dphildBl
        dphildBl->dBldT

        dKdT->dKdphiv[topath="bend right"]
        dKdphiv->dphivdzv
        dphivdzv->dzvdAv
        dzvdAv->dAvdT

        dphivdzv->dzvdBv
        dzvdBv->dBvdT
        dKdphiv->dphivdAv[topath="bend right"]
        dphivdAv->dAvdT
        dKdphiv->dphivdBv
        dphivdBv->dBvdT
        }
    \end{dot2tex}
    \caption{Temperature derivative with respect to K}
    \label{figureK-T}
\end{figure}
The impression of pressure on the phase equilibrium constant K is more pronounced relative to temperature, it influencs the compression factor and parameters A in the equation of the fugacity coefficient, figure(\ref{figureK-P}) illustrates its computational dependence.
\begin{figure}[H]
    \centering
    \begin{dot2tex}[dot,options=-tmath,graphstyle={scale=0.6,transform shape}]
        digraph G {
        node [shape="record"];
        // edge [lblstyle="auto",topath="bend left"];
        edge [lblstyle="auto"];
        dKdphiv [label="\\frac{\\partial K}{\\partial \\phi_{vap}}"]
        dphivdzv [label="\\frac{\\partial \\phi_{vap}}{\\partial z_v}"]
        dphivdAv [label="\\frac{\\partial \\phi_{vap}}{\\partial A_v}"]

        dKdphil [label="\\frac{\\partial K}{\\partial \\phi_{liq}}"]
        dphildzl [label="\\frac{\\partial \\phi_{liq}}{\\partial z_l}"]
        dphildAl [label="\\frac{\\partial \\phi_{liq}}{\\partial A_l}"]

        dzldAl [label="\\frac{\\partial z_l}{\\partial A_l}"]
        dAldP [label="\\frac{\\partial A_l}{\\partial P}"]
        dKdP [label="\\frac{\\partial K}{\\partial P}",style="state, initial"]

        dzvdAv [label="\\frac{\\partial z_v}{\\partial A_v}"]
        dAvdP [label="\\frac{\\partial A_{l}}{\\partial P}"]

        dKdP->dKdphil [topath="bend left"]
        dKdphil->dphildzl
        dphildzl->dzldAl
        dzldAl->dAldP

        dKdphil->dphildAl
        dphildAl->dAldP

        dKdP->dKdphiv[topath="bend right"]
        dKdphiv->dphivdzv
        dphivdzv->dzvdAv
        dzvdAv->dAvdP

        dKdphiv->dphivdAv
        dphivdAv->dAvdP
        }
    \end{dot2tex}
    \caption{Pressure derivative with respect to K}
    \label{figureK-P}
\end{figure}
As with temperature, the effect of molar composition on the phase equilibrium constant K is mediated by acting on the compression factor and parameters A, B in the equation of the fugacity coefficient, figure(\ref{figureK-frac}) illustrates its computational dependence.
\begin{figure}[H]
    \centering
    \begin{dot2tex}[dot,options=-tmath,graphstyle={scale=0.6,transform shape}]
        digraph G {
        node [shape="record"];
        // edge [lblstyle="auto",topath="bend left"];
        edge [lblstyle="auto"];
        dKdphiv [label="\\frac{\\partial K}{\\partial \\phi_{vap}}"]
        dphivdzv [label="\\frac{\\partial \\phi_{vap}}{\\partial z_v}"]
        dphivdAv [label="\\frac{\\partial \\phi_{vap}}{\\partial A_v}"]
        dphivdBv [label="\\frac{\\partial \\phi_{vap}}{\\partial B_v}"]

        dKdphil [label="\\frac{\\partial K}{\\partial \\phi_{liq}}"]
        dphildzl [label="\\frac{\\partial \\phi_{liq}}{\\partial z_l}"]
        dphildAl [label="\\frac{\\partial \\phi_{liq}}{\\partial A_l}"]
        dphildBl [label="\\frac{\\partial \\phi_{liq}}{\\partial B_l}"]

        dzldAl [label="\\frac{\\partial z_l}{\\partial A_l}"]
        dzldBl [label="\\frac{\\partial z_l}{\\partial B_l}"]
        dAldl [label="\\frac{\\partial A_l}{\\partial L_{frac}}"]
        dBldl [label="\\frac{\\partial B_l}{\\partial L_{frac}}"]
        dKdl [label="\\frac{\\partial K}{\\partial L_{frac}}",style="state, initial"]
        dKdv [label="\\frac{\\partial K}{\\partial V_{frac}}",style="state, initial"]

        dzvdAv [label="\\frac{\\partial z_v}{\\partial A_v}"]
        dzvdBv [label="\\frac{\\partial z_v}{\\partial B_v}"]
        dAvdv [label="\\frac{\\partial A_{l}}{\\partial V_{frac}}"]
        dBvdv [label="\\frac{\\partial B_{l}}{\\partial V_{frac}}"]

        dKdl->dKdphil
        dKdphil->dphildzl
        dphildzl->dzldAl
        dzldAl->dAldl

        dphildzl->dzldBl
        dzldBl->dBldl

        dKdphil->dphildAl
        dphildAl->dAldl
        dKdphil->dphildBl
        dphildBl->dBldl

        dKdv->dKdphiv
        dKdphiv->dphivdzv
        dphivdzv->dzvdAv
        dzvdAv->dAvdv

        dphivdzv->dzvdBv
        dzvdBv->dBvdv
        dKdphiv->dphivdAv
        dphivdAv->dAvdv
        dKdphiv->dphivdBv
        dphivdBv->dBvdv
        }
    \end{dot2tex}
    \caption{Mole composition derivative with respect to K}
    \label{figureK-frac}
\end{figure}
In the above derivative chain rule diagram, all of them are explicit expressions except for the derivatives of the compression coefficient with respect to parameters A, B. In the calculation of the exact derivative of the compression factor with respect to the parameters, the roots of the cubic equation are calculated by the discriminant, and the traditional automatic differentiation method cannot satisfy the automatic differentiation derivative calculation for the control logic, while the method based on the source code transformation used in this paper can be better expanded by the conditional logic judgment and can generate the exact derivative at any position. \par
\begin{figure}[H]
    \centering
    \subfigure[Methane]{
        \label{figuredKdT.sub.1}
        \includegraphics[width=0.45\textwidth]{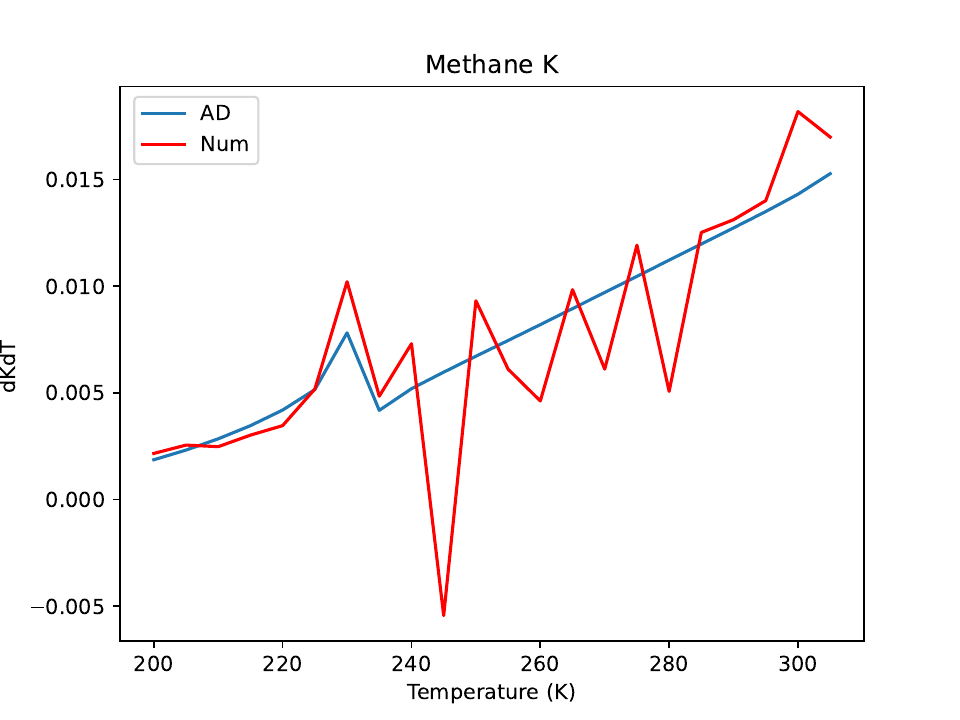}}
    \subfigure[Ethane]{
        \label{figuredKdT.sub.2}
        \includegraphics[width=0.45\textwidth]{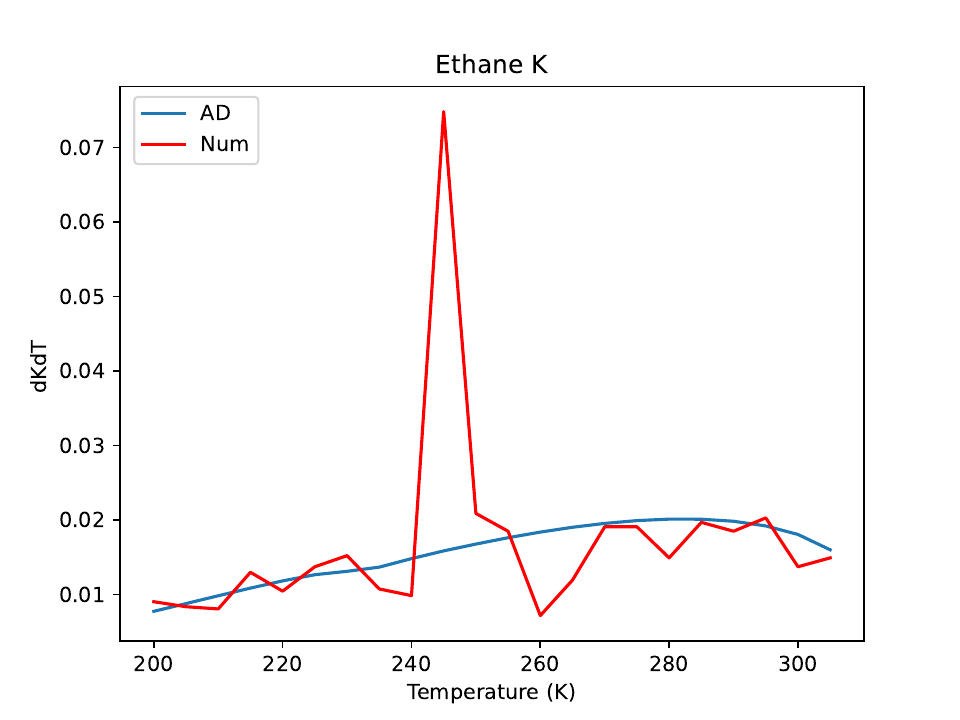}}
    \subfigure[Ethylene]{
        \label{figuredKdT.sub.3}
        \includegraphics[width=0.45\textwidth]{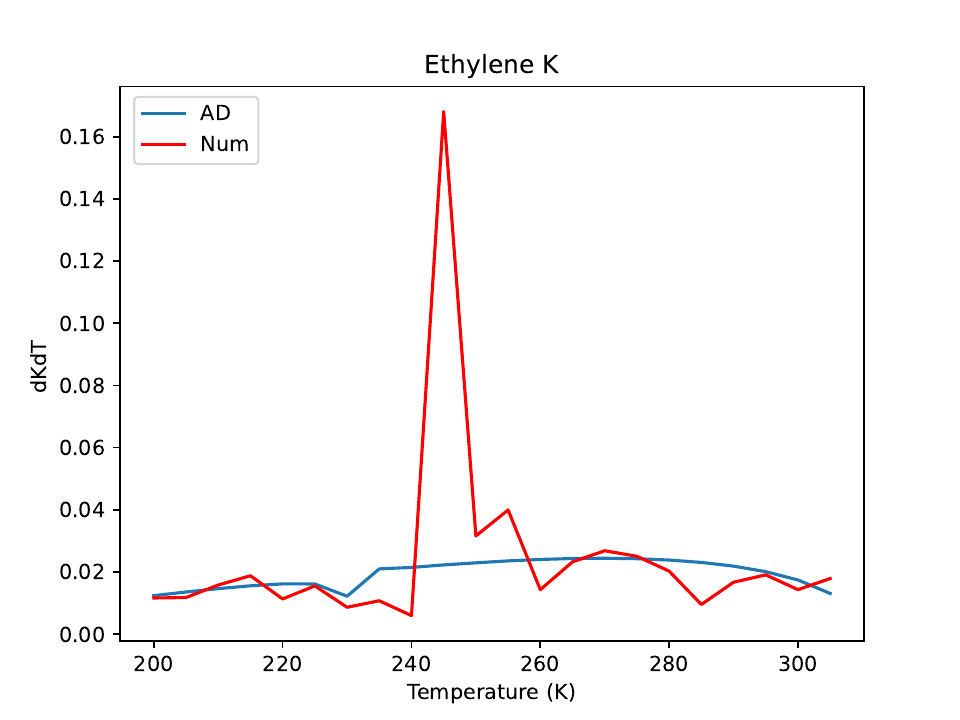}}
    \subfigure[Propane]{
        \label{figuredKdT.sub.4}
        \includegraphics[width=0.45\textwidth]{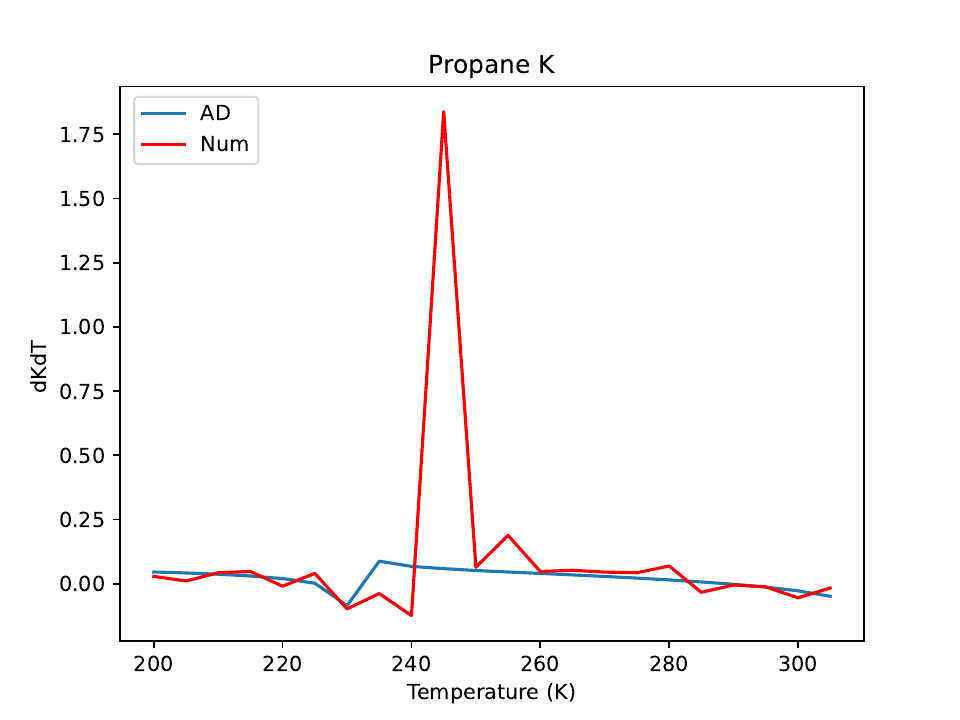}}
    % \subfigure[Propylene]{
    %     \label{figuredKdT.sub.5}
    %     \includegraphics[width=0.45\textwidth]{figure/dkd/dKdTC3H6.pdf}}
    \caption{The derivatives of K with respect to temperature for different substances}
    \label{figuredKdT}
\end{figure}

At a molar composition of methane ethane ethylene propane (0.25, 0.25, 0.25, 0.25) at a temperature of 200K-300K and a pressure of 18 bar, numerical experiments on temperature derivatives were conducted.
% In this part, numerical experiments on the derivatives of phase equilibrium constants with respect to temperature are conducted using a number of significant ethylene production (methane, ethane, ethylene, propane) components. 
Figure (\ref{figuredKdT}) shows the derivatives of the phase equilibrium constants K with respect to temperature for different substances obtained using the automatic differentiation method and the numerical differentiation method, and it can be seen that the results obtained by the automatic differentiation method are significantly smoother and more stable. 

\par
Due to the existence of truncation error in the numerical approach, the derivative curve swings greatly; the auto-differentiation method does not exhibit considerable jitter in the overall curve since there is no truncation error in the calculation process. The numerical method exhibits an obvious jump between 240 and 260K due to the change in the discriminant of the cubic equation for calculating the compression factor, the different formula for calculating the compression factor, and the accumulation of numerical differentiation errors, whereas the automatic differentiation can generate the analytic derivative accurately at this point, so the transition can be smooth.\par

\begin{figure}[H]
    \centering
    \subfigure[Methane]{
        \label{figuredKdP.sub.1}
        \includegraphics[width=0.45\textwidth]{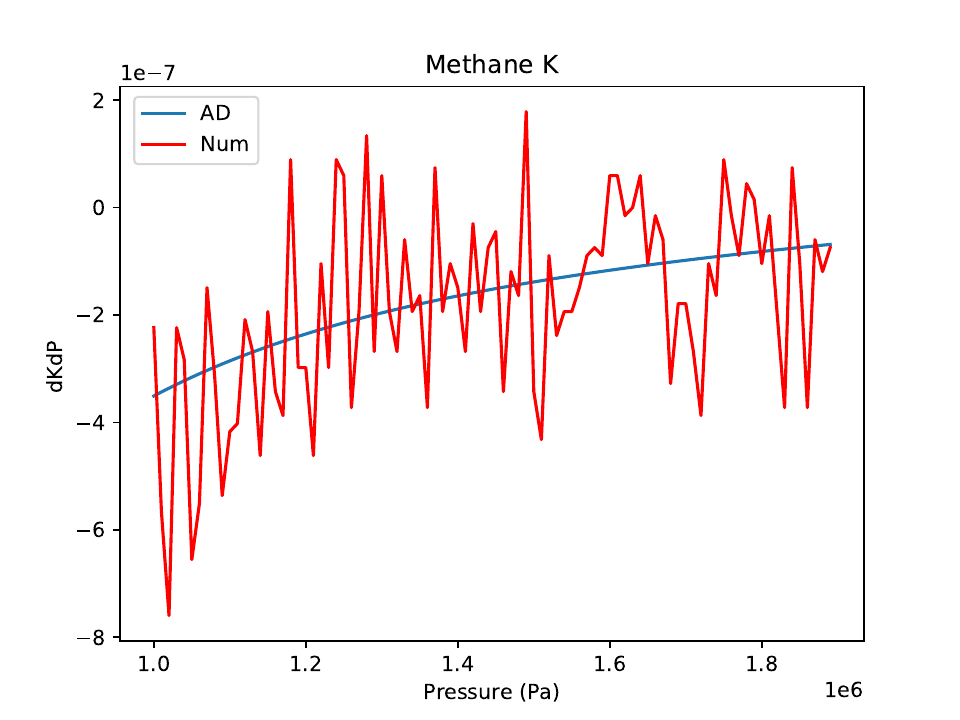}}
    \subfigure[Ethane]{
        \label{figuredKdP.sub.2}
        \includegraphics[width=0.45\textwidth]{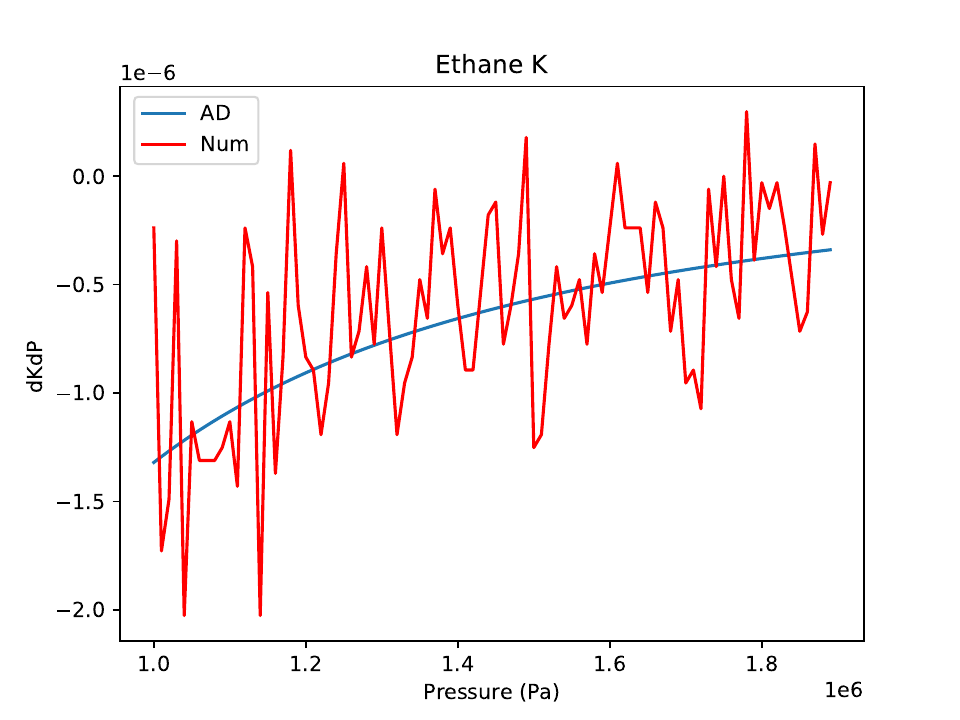}}
    \subfigure[Ethylene]{
        \label{figuredKdP.sub.3}
        \includegraphics[width=0.45\textwidth]{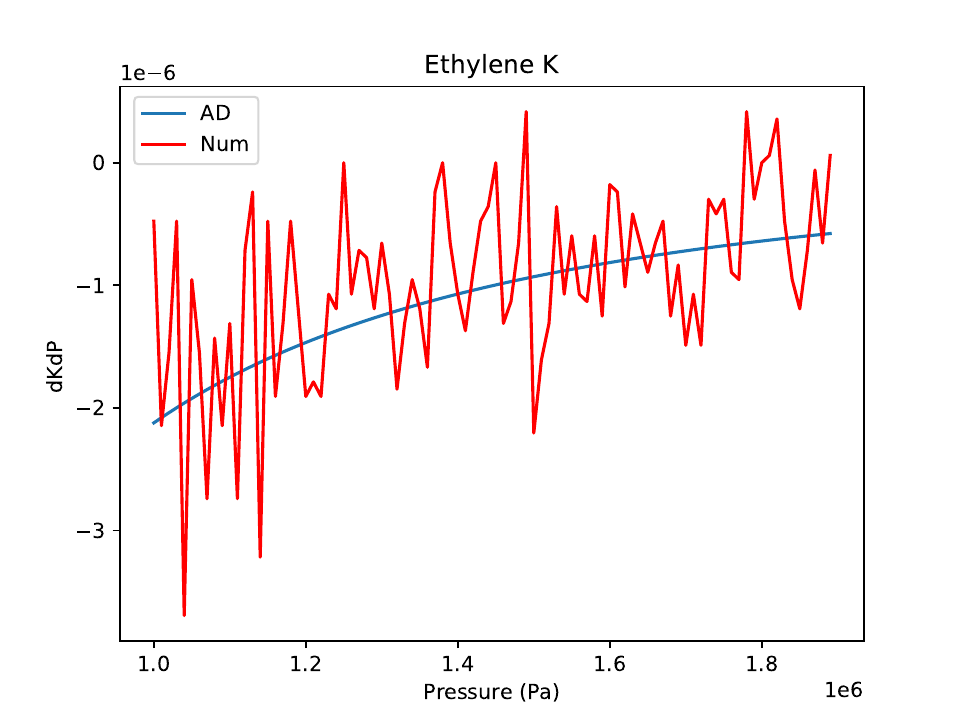}}
    \subfigure[Propane]{
        \label{figuredKdP.sub.4}
        \includegraphics[width=0.45\textwidth]{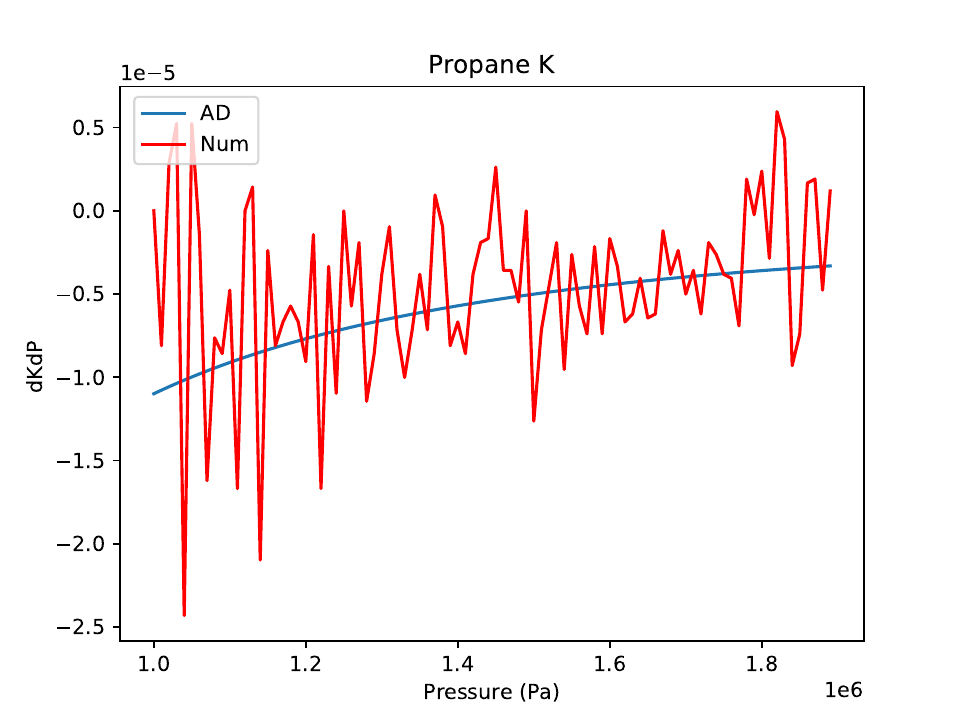}}
    % \subfigure[Propylene]{
    %     \label{Fig6.sub.5}
    %     \includegraphics[width=0.45\textwidth]{figure/dkd/dKdTC3H6.pdf}}
    \caption{The derivatives of K with respect to temperature for different substances}
    \label{figuredKdP}
\end{figure}

\begin{figure}[H]
    \centering
    \subfigure[step = 50]{
        \label{Fig9.sub.1}
        \includegraphics[width=0.3\textwidth]{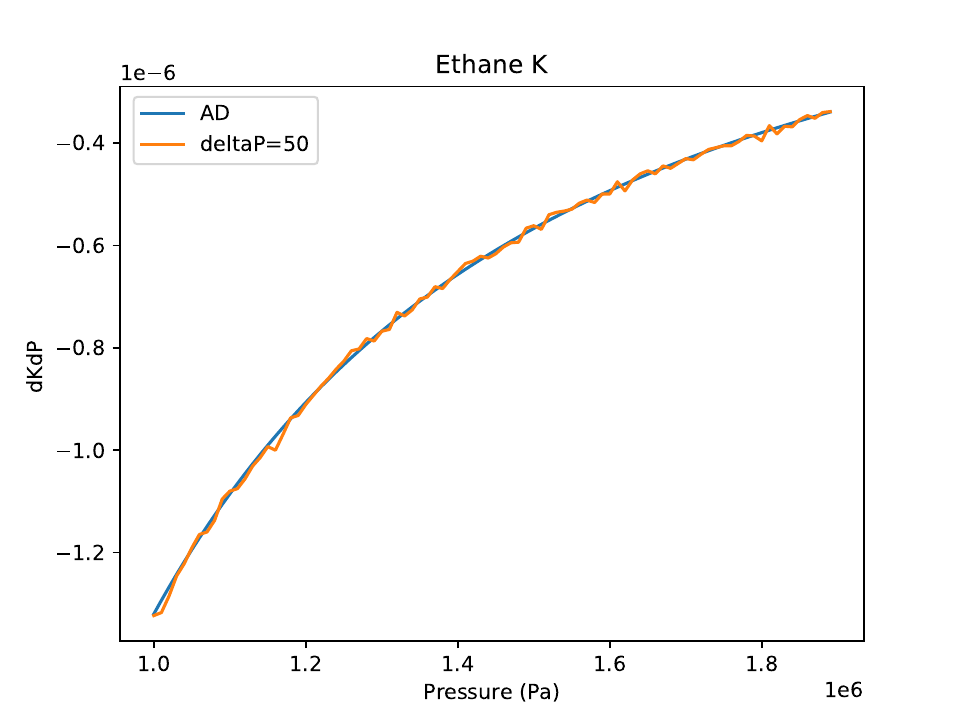}}
    \subfigure[step = 10, 5]{
        \label{Fig9.sub.2}
        \includegraphics[width=0.3\textwidth]{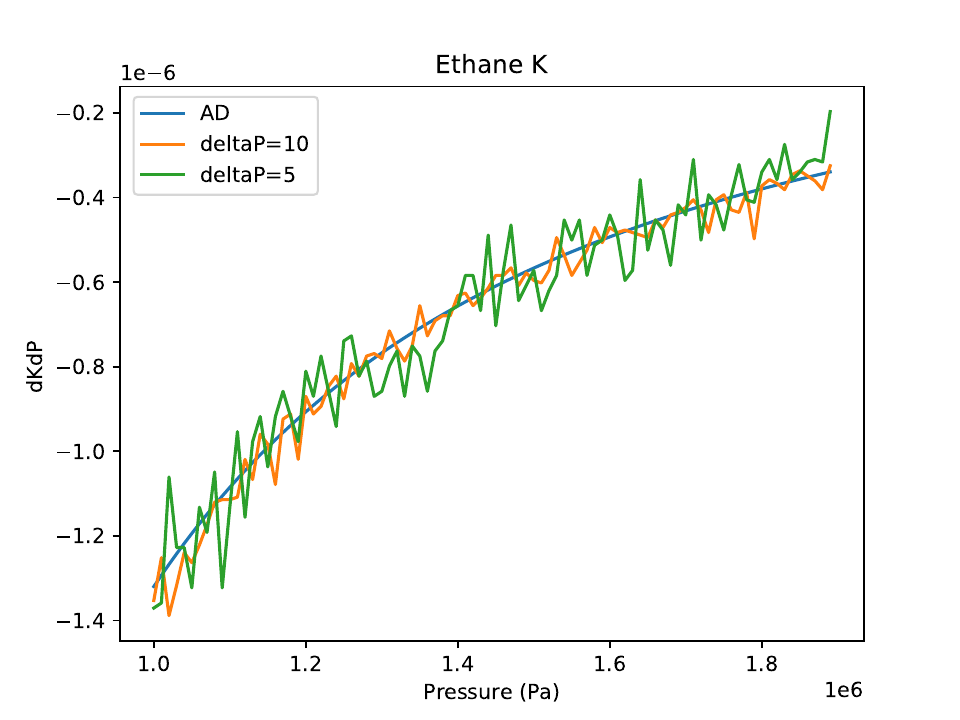}}
    \subfigure[step = 1, 0.1]{
        \label{Fig9.sub.3}
        \includegraphics[width=0.3\textwidth]{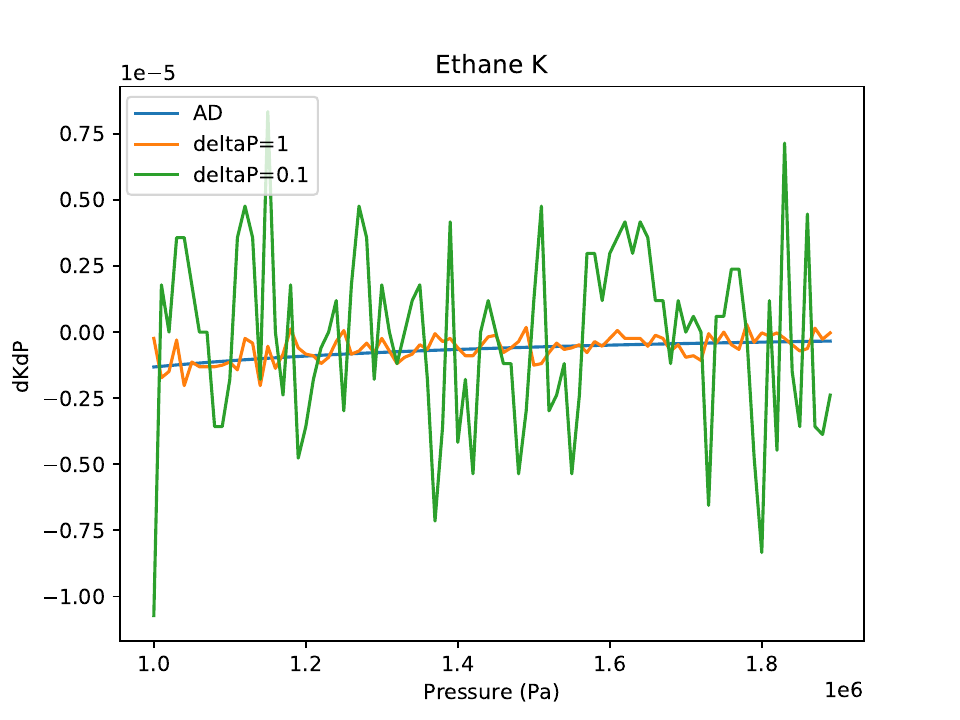}}
    \caption{Different differentiation step of numerical method}
    \label{figuredKdP_delta}
\end{figure}

Figure (\ref{figuredKdP})  illustrates the derivatives of the phase equilibrium constants K with respect  to the pressure for different substances through the automatic differentiation method and the numerical differentiation method. Similar to the results for temperature. The results for pressure obtained by the automatic differentiation method are significantly smoother and more stable. With the pressure changing, the numerical differentiation fluctuates dramatically. This is because the derivative of the finite difference calculation can be affected obviously by the computer floating point round-off error with the pressure as high as 10 bar-19 bar.
Furthermore, the finite difference method's difference step significantly affects the results, especially when there are significant fluctuations in the variables being simulated. The reliability of the derivative results could be compromised if the difference step can't be adjusted adaptively. Figure(\ref{figuredKdP_delta}) demonstrates the contrast between numerical derivatives and auto-differentiation outcomes with multiple step sizes. Theoretically, a smaller step size would yield more precise differentiation outcomes. Nevertheless, when absolute values are significant, meaningful accuracy is reduced by defects in the computational process of floating-point calculations. Notably, decreasing the differential step size leads to a considerable surge in error.
In the past studies, researchers have tried to improve the stability of numerical difference by modifying the finite difference format\cite{kamath2010equation} and dynamically adjusting the iteration step\cite{ErrorAnalysis}, etc, which can significantly improve the convergency and efficiency.However, the problem solved in the publications are non-typical and difficult to be applied to other computational models. The auto-differential method is effective in improving both computational stability and highly scalable, which will bedemonstrated in flash calculations on the basis of differentiable thermodynamics in the following section.
% \begin{figure}[H] 
%     \centering 
%     \includegraphics[width=0.5\textwidth]{H2-C3_CalcK.pdf} 
%     \caption{K and derivatives of K} 
%     \label{figure5} 
% \end{figure}
\subsection{Flash calculations}
% Flash calculation is the core of almost all process simulation calculations. An accurate and efficient flash algorithm directly determines the efficiency of the full process calculation optimization. The nested loops inner and outer nesting algorithm is a stable and effective method for flash calculation, which solves the phase equilibrium fraction in the inner layer while the outer layer performs the heat balance calculation. This section will show the implementation of the nested loops algorithm using automatic differentiation, and demonstrate the convergence improvement of flash calculation using automatic differentiation algorithm.
Almost all process simulation calculations center on flash calculation. An accurate and efficient flash algorithm directly impacts the effectiveness of the optimization of the entire process calculation. The nested loops inner and outer nesting algorithm is a reliable and efficient method for flash computation, which solves the phase equilibrium fraction in the inner layer and the heat balance calculation in the outer layer. This section will showcase the implementation of the nested loops algorithm utilizing automatic differentiation, as well as the improvement in convergence of the flash calculation algorithm utilizing automatic differentiation.\par
% and the parallel computation of large-scale vectors by splitting the thermodynamic computation graph to achieve chemical simulation operator level parallelism in the face of large-scale optimization.

\begin{figure}[H] 
    \centering 
    \includegraphics[width=0.5\textwidth]{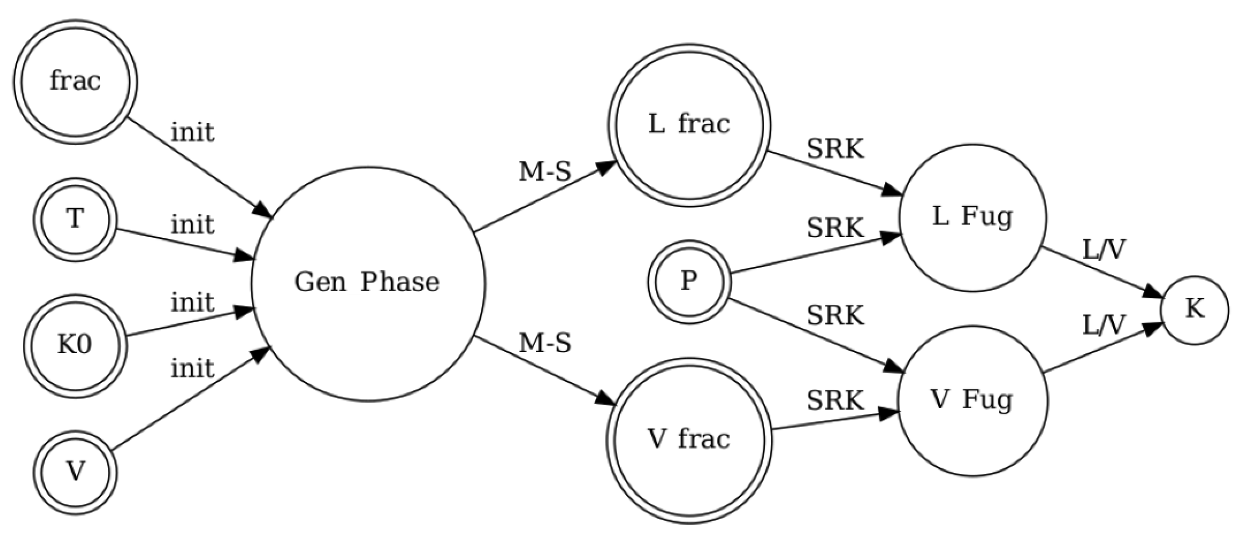} 
    \caption{Calculation of K in flash algorithm}
    \label{figureCalcKinflash} 
\end{figure}

Flash  calculation is based on material balance Eq.(\ref{FVL}-\ref{fvl}), mole-fraction summations Eq.(\ref{xi}-\ref{yi}), phase equilibrium Eq.(\ref{kxy}) and energy conservation Eq.(\ref{Hxy}).\par
\begin{gather}
    F = V+L \label{FVL}\\
    f_i = L*x_i+V*y_i\label{fvl}\\
    \sum{x_i} =1 \label{xi}\\
    \sum{y_i} =1 \label{yi}\\
    K_i = \frac{y_i}{x_i} \label{kxy}\\
    H_f*F+Q = H_L*L+H_V*V \label{Hxy}
\end{gather}
Despite the fact that individual flash calculations have distinct design specifications, the objective is to answer the aforementioned conservation relation equation by changing the design variables to the feed condition.
The flash algorithm process consists of \textit{Phase generation}, \textit{Fugacity calculate}, \textit{Components summation}, \textit{Thermodynamic equilibrium}. Figure({\ref{figureCalcKinflash}}) displays the straightforward computational graph for flash calculation.  Eq.(\ref{xl}-\ref{yv}) use vapor fraction, mixture phase fraction and K calculate the liquid and vapor fraction, these equations are the pahse generation function.
After phase generation, Eq.(\ref{SRKfunc}-\ref{phi_vap}) use SRK state equation calculate the fugacity coefficient of liquid and vapor phase,The differentiable thermodynamic approach discussed in the previous section is mainly useful here. Components summation Eq.(\ref{ObjectFun})is the inner iteration convergence equation, When solved using the iterative method, the derivatives of the convergence equation will be obtained using automatic differentiation.\par
\begin{gather}
    x_l = \frac{z_{mix}}{(K-1)*V+1} \label{xl}\\
    y_v = \frac{z_{mix}K}{(K-1)*V+1}\label{yv}\\
    F = \sum_{i}(y_{vi}-x_{li})\label{ObjectFun}
\end{gather}

% \begin{figure}[H]
%     \centering
%     \subfigure[CalcK in flash]{
%         \label{Fig6.sub.1}
%         \includegraphics[width=0.45\textwidth]{Flash_K.png}}
%     \subfigure[PV flash]{
%         \label{Fig6.sub.2}
%         \includegraphics[width=0.5\textwidth]{PV_.png}}
%     \caption{Computational Graph of Flash}
%     \label{figure6}
% \end{figure}

This section explains the benefit of the automatic differentiation technique in PT, PV, and PH flash, as well as the convergence feature for nested computations.

\subsubsection{Flash PT }
\begin{figure}[H]
    \centering
    \subfigure[T = 180K]{
        \label{Fig7.sub.1}
        \includegraphics[width=0.45\textwidth]{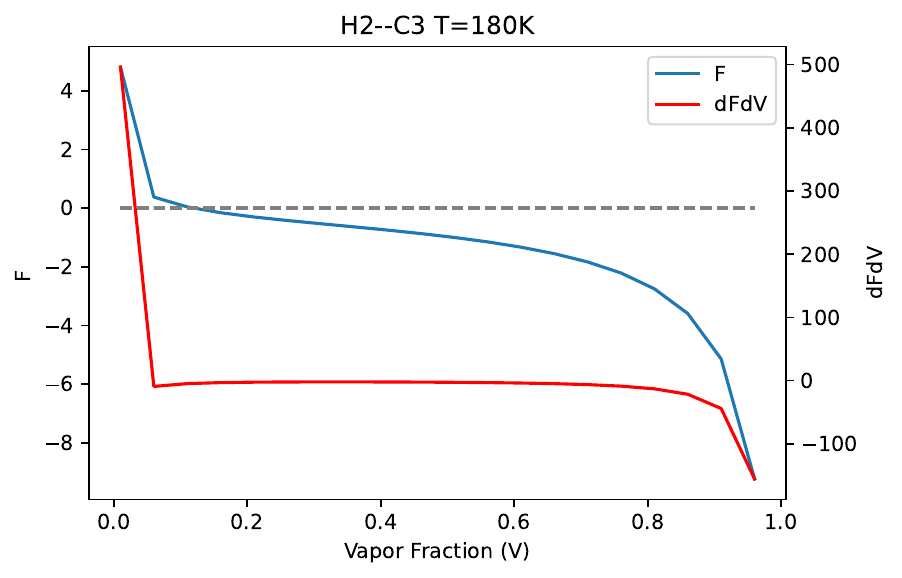}}
    \subfigure[T = 220K]{
        \label{Fig7.sub.2}
        \includegraphics[width=0.45\textwidth]{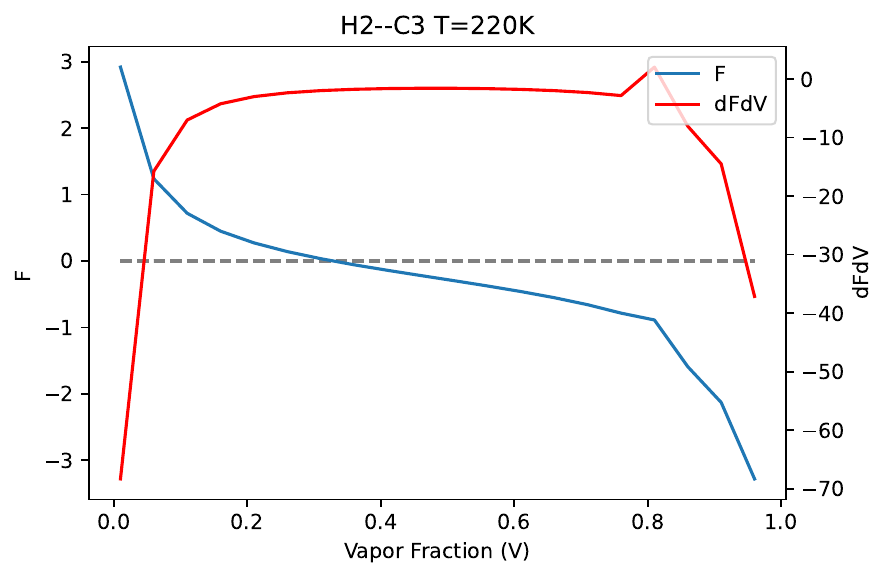}}
    \quad
    \subfigure[T = 240K]{
        \label{Fig7.sub.3}
        \includegraphics[width=0.45\textwidth]{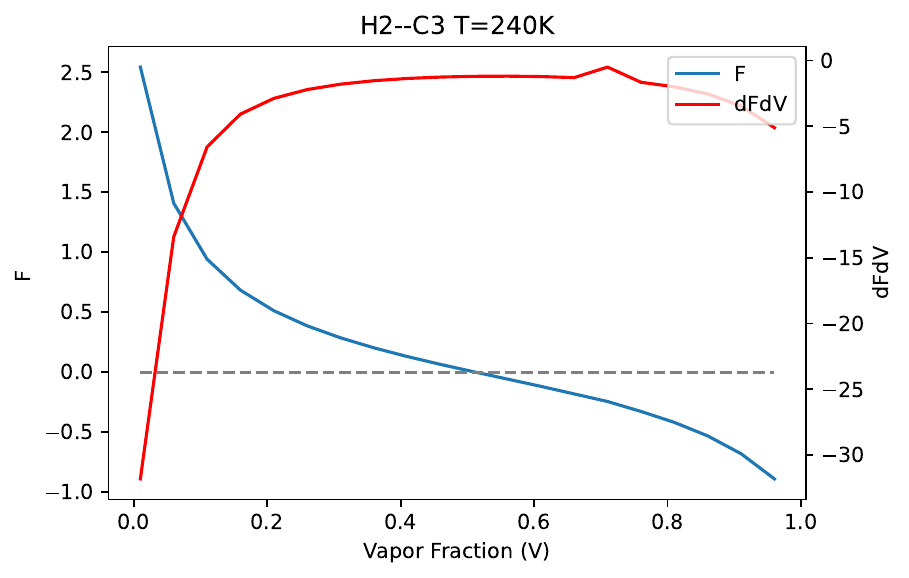}}
    \subfigure[T = 260K]{
        \label{Fig7.sub.4}
        \includegraphics[width=0.45\textwidth]{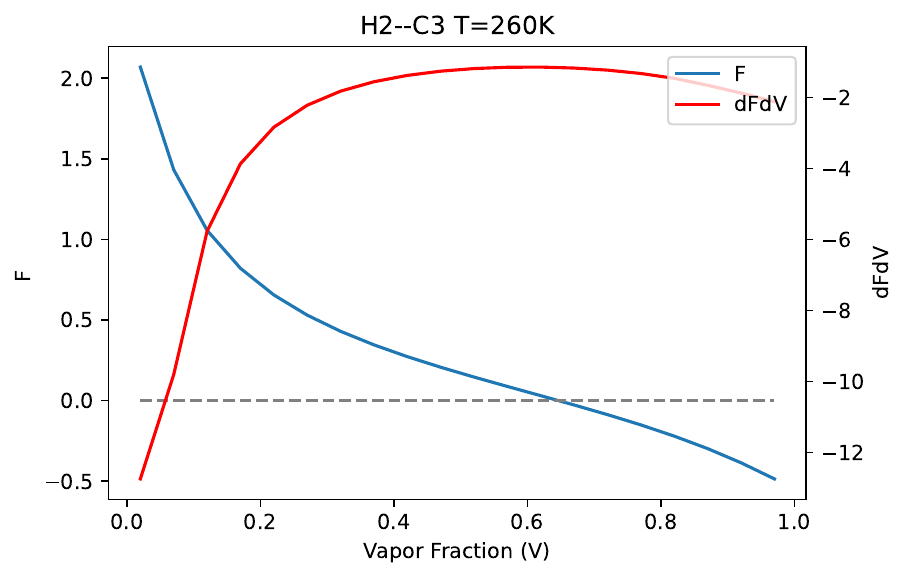}}
    \caption{Convergence objective F and the derivative of F with respect to V}
    \label{figurePT}
\end{figure}

In pressure and temperature flash, the convergence equation can be translated into Eq(\ref{objective}) since the condition provides the temperature and pressure, and the unknown variable is the component gas phase fraction. After phase determination the calculation will iterate V (vapor fraction). The iteration is usually performed using Newton's method, automatic differentiation thermodynamic provide the derivative of K with respect to V.\par
\begin{gather}
    F(V) = \sum_{i}\frac{z_{mix}(K_i-1)}{(K_i-1)*V+1}\label{objective}
\end{gather}
% Pressure and temperature flash is relatively easy, because the convergence equation is explicitly, the figure({\ref{figure7}}) illustrates the values of the convergence equations of those substances with the derivatives at different temperatures.
Since the parameter V affects the convergence equation explicitly, its derivative may be directly determined using the formula technique. Figure(\ref{figurePT}) displays the convergence equation values derived with their derivatives for various V under various temperature settings. The intersection of the dotted line with the F curve denotes the state where the convergence equation at this point is equal to 0, i.e., the computational convergence.

\subsubsection{Flash PV }
\begin{figure}[H]
    \centering
    \subfigure[Vapor fraction = 0.7 AD]{
        \label{figurePV.sub.1}
        \includegraphics[width=0.45\textwidth]{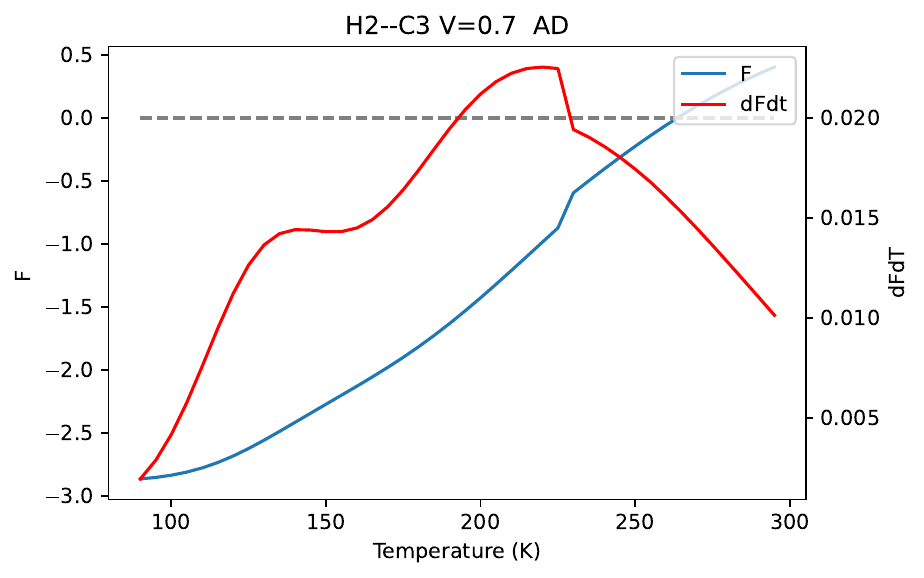}}
    \subfigure[Vapor fraction = 0.7 numerical]{
        \label{figurePV.sub.2}
        \includegraphics[width=0.45\textwidth]{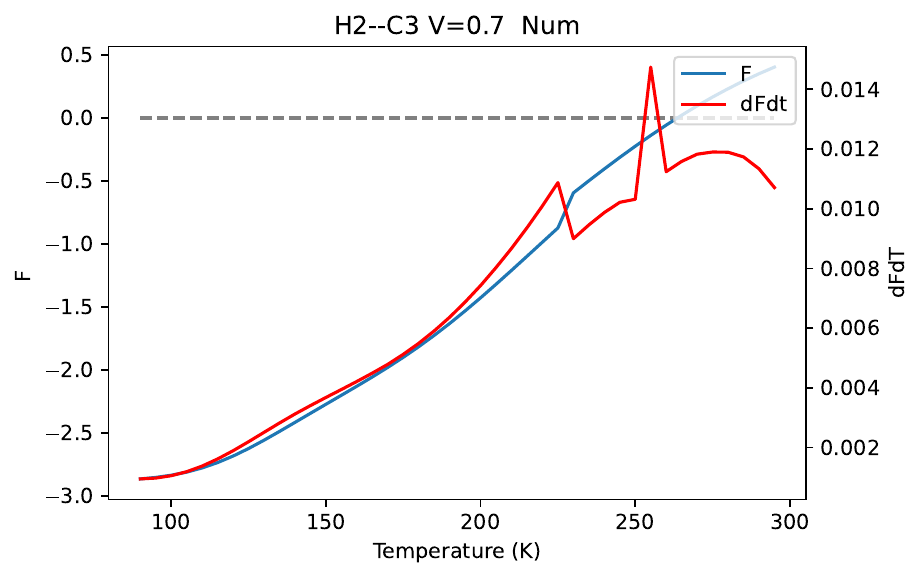}}
    \quad
    \subfigure[Vapor fraction = 0.9 AD]{
        \label{figurePV.sub.3}
        \includegraphics[width=0.45\textwidth]{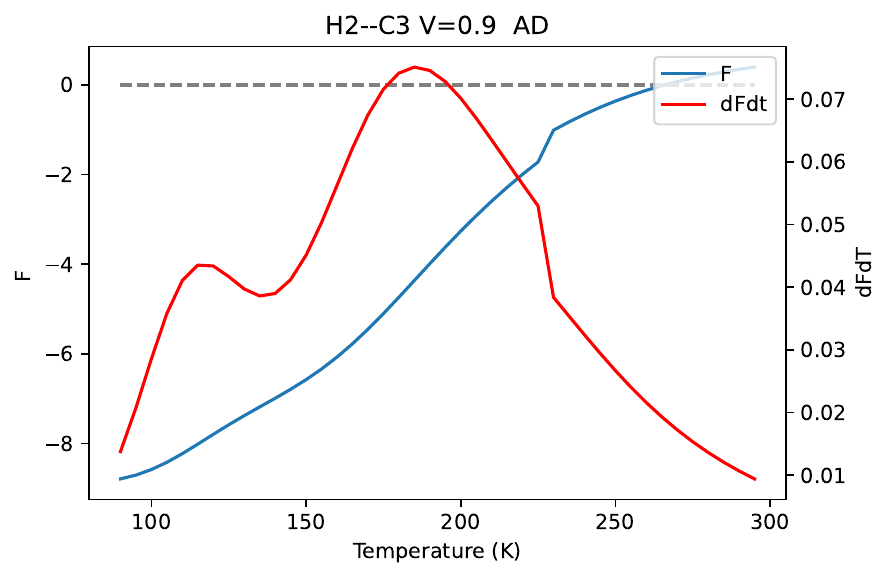}}
    \subfigure[Vapor fraction = 0.9 Numerical]{
        \label{figurePV.sub.4}
        \includegraphics[width=0.465\textwidth]{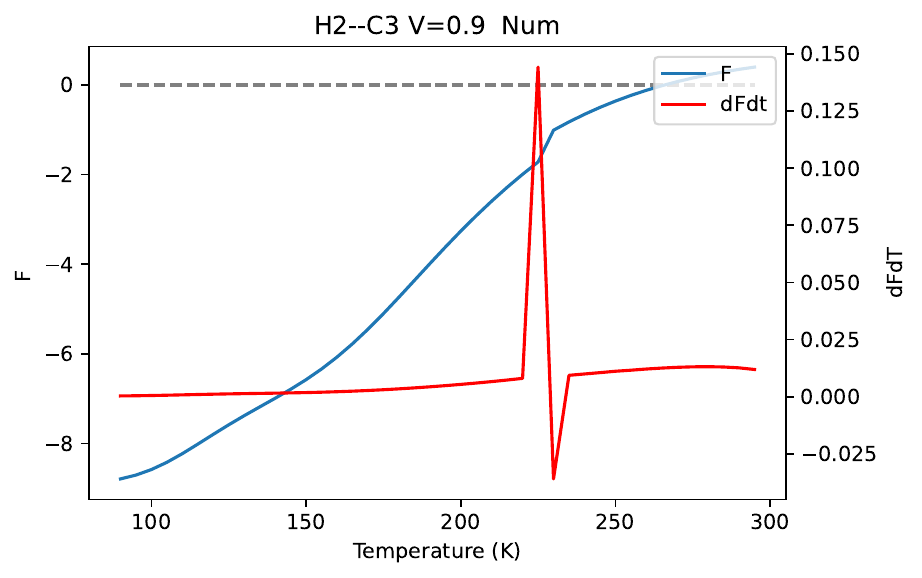 }}
    \caption{Convergence objective F and the derivative of F with respect to T}
    \label{figurePV}
\end{figure}

For flash PV, the pressure and vapor fraction flash has the vapor fraction that has been set, temperature is the iterate variable, but temperature is implicitly related to the convergence objective F. Traditional way to get the derivative of F with respect to temperature is using numerical differentiation, Eqs.(\ref{objective_T}-\ref{dKdT}) are the numerical differentiation for the flash convergence objective.\par
\begin{gather}
    F(T) = \sum_{i}\frac{z_{mix}(K_i(T)-1)}{(K_i(T)-1)*V+1}\label{objective_T}\\
    \frac{dF}{dT} = \sum_{i}\frac{dF}{dK}*\frac{dK}{dT}\\
    \frac{dF}{dK} = \frac{z_{mix}}{(K_T-1)*V+1}\\
    \frac{dK}{dT} = \frac{K_{T+\Delta T}-K_{T-\Delta T}}{2\Delta T}\label{dKdT}
\end{gather}
In PV calculation, Newton's iterative method is the commonly used algorithm. Eqs.(\ref{newton}) is the basic form of Newton's iterative formula, where $\frac{dF}{dT}$ is the most important value for the calculation, figure(\ref{figurePV}) shows the difference between AD and numerical differentiation, It is found that AD can obtain the smoother derivatives as compared to the numerical differentiation, which even undergoes mistake during the derivation because of the error accumulation derived from inappropriate differential step size.\par
\begin{gather}
    T_{i+1} = T_{i} - \frac{F(T_i)}{\frac{dF}{dT_i}}\label{newton}
\end{gather}
The errors of the numerical method exhibit two cases of slow computational convergence and computational convergence failure as shown in Figure (\ref{figurePV}) . In Figure (\ref{figurePV}b), the absolute value of the derivative with respect to temperature is small and the theoretical, the Newton's(\ref{newton}) method implemented using numerical methods does not achieve the theoretical rate of convergence of the Newton's method because $dx$ is not accurate and the values are large. In addition to the small absolute value of the inverse, there is also a case of numerical differentiation error, at which the Newton iteration is misdirected, and therefore lead to  divergence or oscillation of convergence.\par
\begin{figure}[H]
    \centering
    \subfigure[step = 0.1,0.01,0.001]{
        \label{figurePVstep_error.sub.1}
        \includegraphics[width=0.45\textwidth]{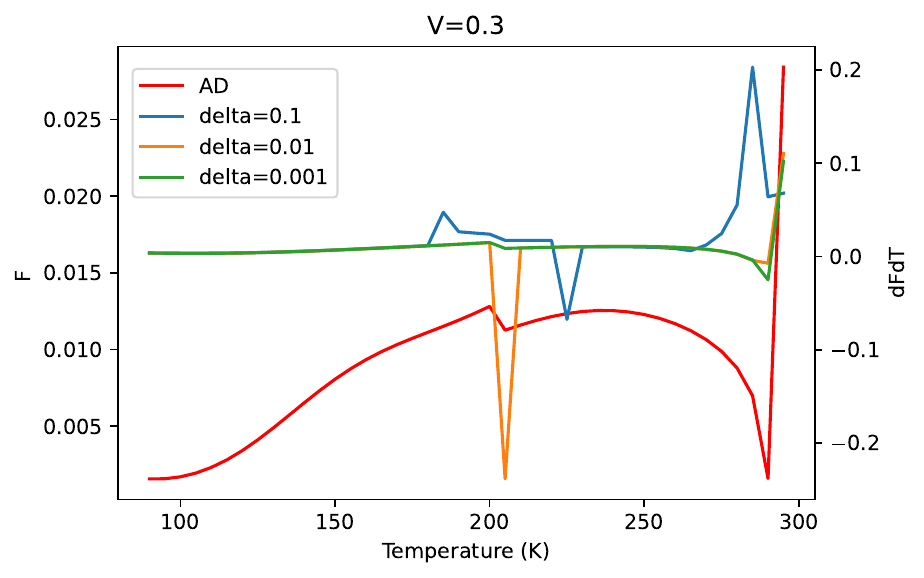}}
    \subfigure[step=0.0001,1e-5,1e-6]{
        \label{figurePVstep_error.sub.2}
        \includegraphics[width=0.45\textwidth]{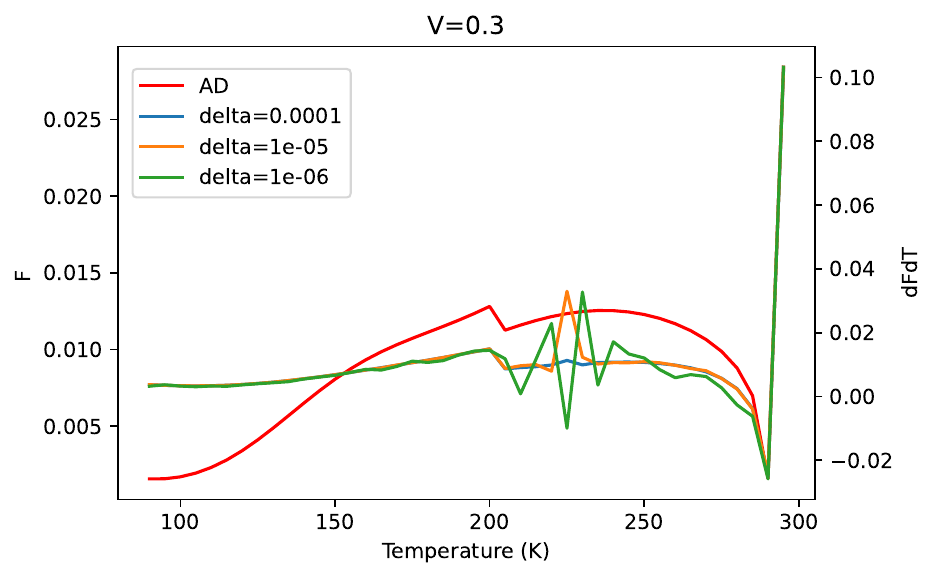}}
    \quad
    \subfigure[step=1e-7]{
        \label{figurePVstep_error.sub.3}
        \includegraphics[width=0.45\textwidth]{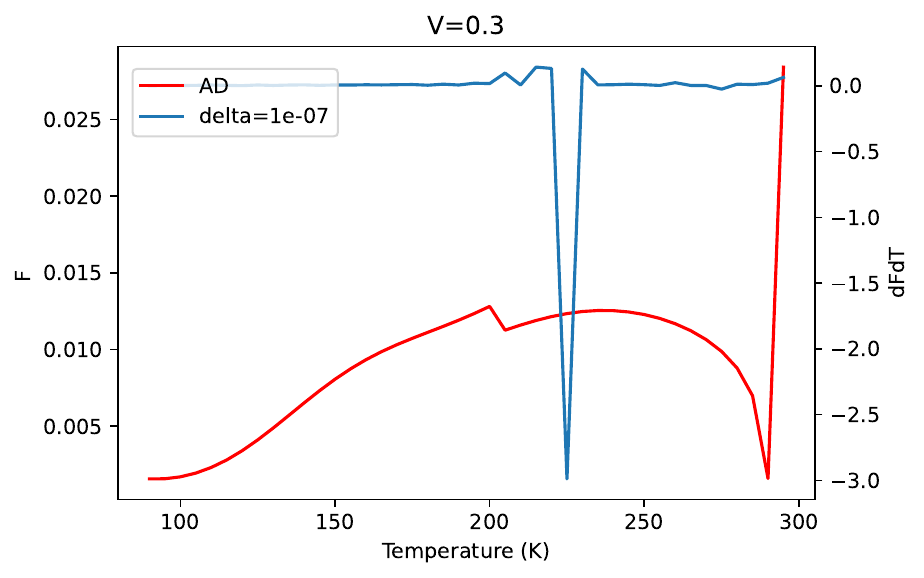}}
    \subfigure[step=1e-8]{
        \label{figurePVstep_error.sub.4}
        \includegraphics[width=0.45\textwidth]{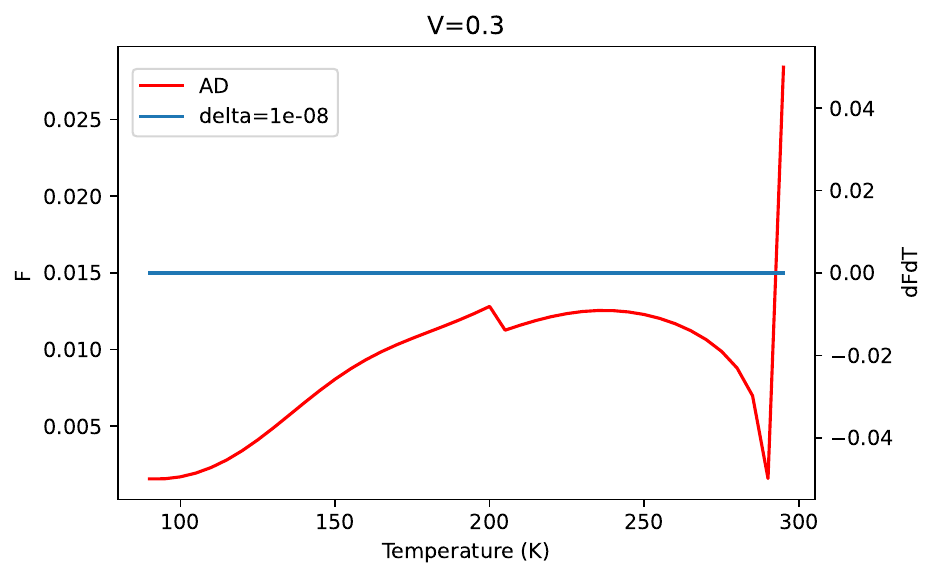}}
    \caption{Different differentiation step of numerical method }
    \label{figurePVstep_error}
\end{figure}

In process simulation calculation, the differential step size should be adaptively adjusted by a program to prevent error accumulation. However, the mechanisms of robustness and applicability is difficult to verify. Figure(\ref{figurePVstep_error}) illustrates the effect of various differentiation steps on the derivative values in detail. The derivatives are not smooth all the steps. With the iteration step increases, the calculation results  diverge from the exact results. While with the iteration step reduced to 1e-6, the numerical derivative results slightly improve but still exhibit obvious fluctuations. If the iteration step is further decreased to 1e-7, there will be obvious errors in the calculation due to the numerical accuracy, which may even lead to an infeasible calculation.\par

\subsubsection{Flash PH }
\begin{figure}[H] 
    \centering 
    \includegraphics[width=0.7\textwidth]{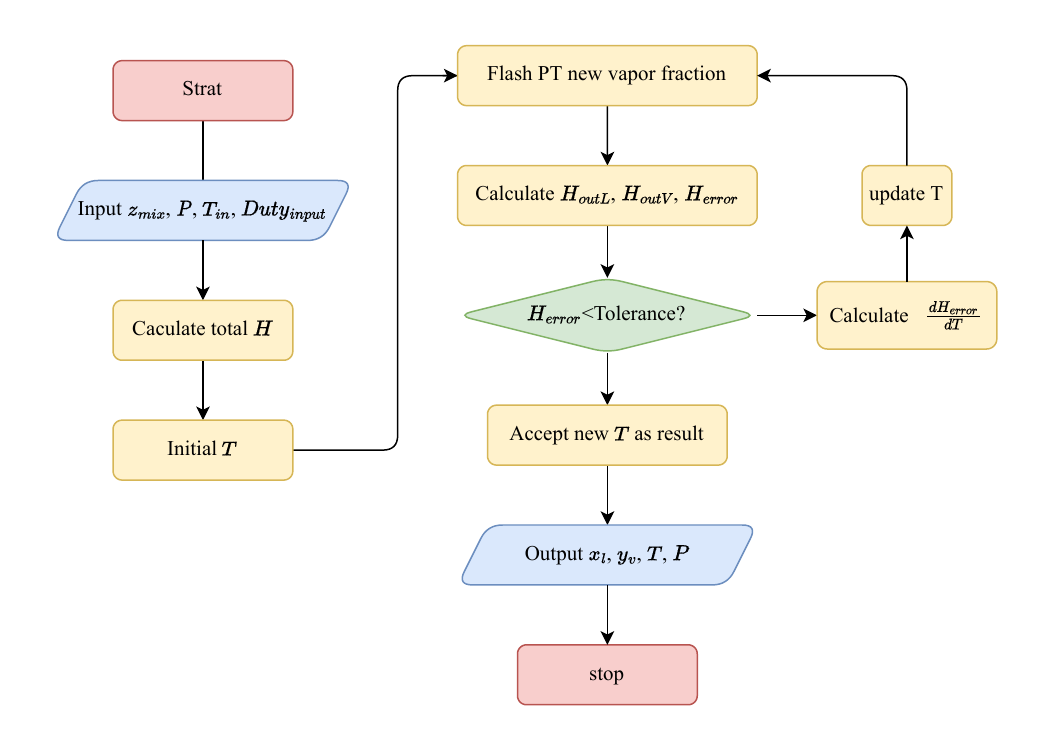} 
    \caption{Algorithm for PH Flash} 
    \label{figure101} 
\end{figure}

Adiabatic flash or Duty flash is the basic for mixing and separation process, the base rule for the calculation is heat balance, as shown in Eq.(\ref{Hbalance}). The $H_{error}$ is the convergence equation, the algorithm will iterate temperature until $H_{error}$ less than Tolerance. The inner loop for the algorithm is flash PT, and the $ H_{outL}, H_{outV}, H_{out} $ will be calculated with vapor fraction, $V_{out}, x_{l}, y_{v}$, obtained from flash PT.\par
\begin{gather}
    H_{total} = H_{mix} + Q= \sum H_{out}\label{Hbalance}\\
    H_{mix} = (1-V)*H_{mixL}+ V*H_{mixV}\\
    H_{out} = (1-V_{out})H_{outL}+V_{out}H_{outV}\\
    H_{error} = H_{total}-H_{out}\\
\end{gather}
The outer loop is the heat balance calculation. In this loop, temperature is implicitly related to the convergence objective $H_{error}$ because the inner loop is flash PT. The traditional method is iterating the temperature interval, which is very inefficient \cite{seader2016separation}
Newton method is a practical way to update temperature, as shown in Eq.(\ref{H_iter}). The flow chart of the algorithm is shown in figure(\ref{figure101}).\par
\begin{gather}
    T_{i+1} = T_{i} - \frac{H_{error}(T)}{\frac{dH_{error}}{dT}}\label{H_iter}\\
    \frac{dH_{error}}{dT} = \frac{H_{error}(T+\delta)-H_{error}(T-\delta)}{2\delta}\\
\end{gather}
In the nested loop flash algorithm,  a precise result for $\frac{dH_{error}}{dT}$ is the key to the success of the algorithm. In traditional numerical differentiation method, flash PT should be calculate firstly to get  the vapor phase fraction, vapor mole fraction and liquid mole fraction, so as to calculaye the enthalpy. The automatic differentiation thermodynamic can obtain the exact derivatives for $H_{error}$.\par
\begin{figure}[H] 
    \centering 
    \subfigure[automatic differentiation method]{
        % \label{Fig.sub.1}
        \includegraphics[width=0.4\textwidth]{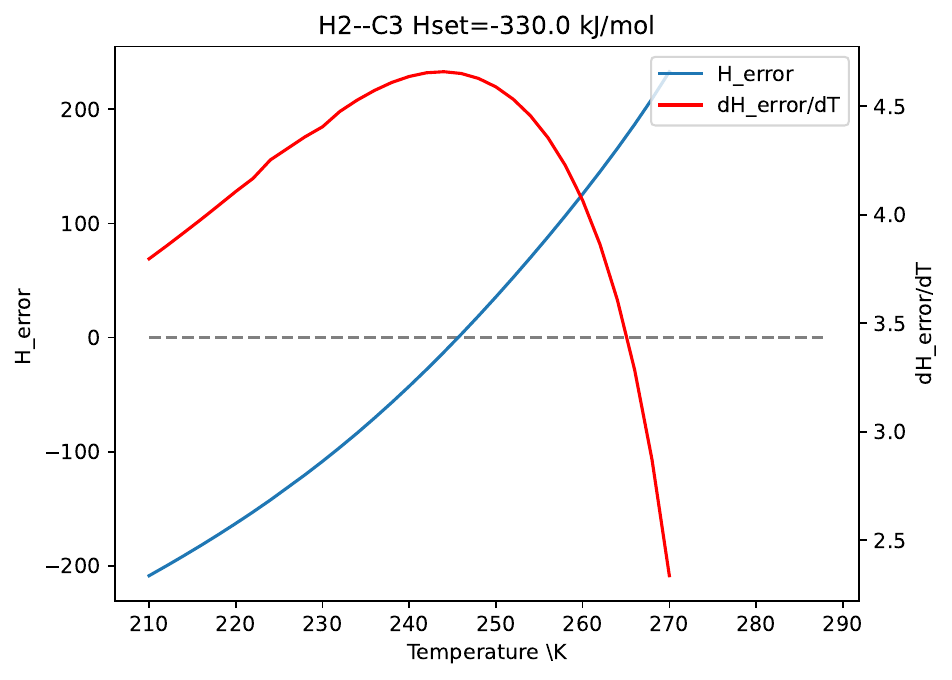}}
    \subfigure[numerical method]{
        % \label{Fig.sub.2}
        \includegraphics[width=0.4\textwidth]{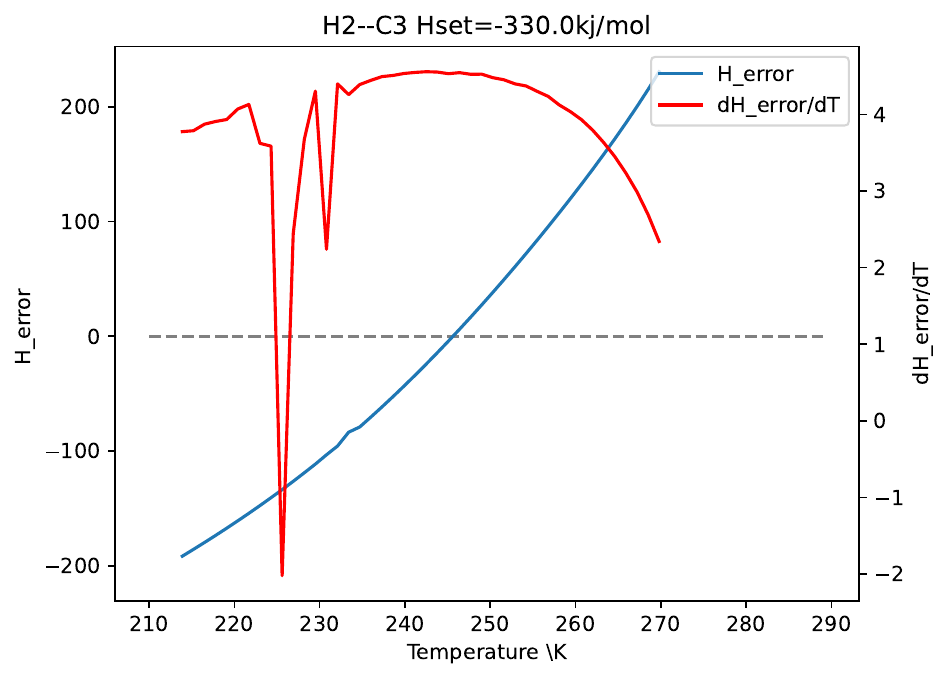}}
    \caption{Two method for $H_{error}$ and $\frac{dH_{error}}{dT}$ } 
    \label{figure12} 
\end{figure}

In figure(\ref{figure12}) , the automatic differentiation method can get the $\frac{dH_{error}}{dT}$ smoother than numerical method, if the initial value is in the wide oscillation interval, the pressure and enthalpy flash will fail to converge. In the PH flash condition, the inner loops contain PT flash and enthalpy calculation, it is difficult to realize the traditional manual derivation method, the automatic differentiation is the most effective and stable method.\par
% \section{Results and discuss}
% \subsection{Results summary}
% In this section, the process simulation framework using automatic differentiation will be used to simulate real industrial processes, and the results will be compared with commercial process simulation software to illustrate the enhancement of process simulation calculations brought by using the automatic differentiation method in terms of convergence, robustness of initial value setting, and explanation of simulation convergence failures. In order to fully demonstrate the superiority brought by the automatic differentiation method, the results of some unit operation calculations in ethylene production are simulated and analyzed using the SRK physical property method.
\subsection{Evaluation of convergence}
\begin{figure}[H]
    \centering
    \subfigure[Vapor fraction = 0.7 AD]{
        \label{figurecompare.sub.1}
        \includegraphics[width=1\textwidth]{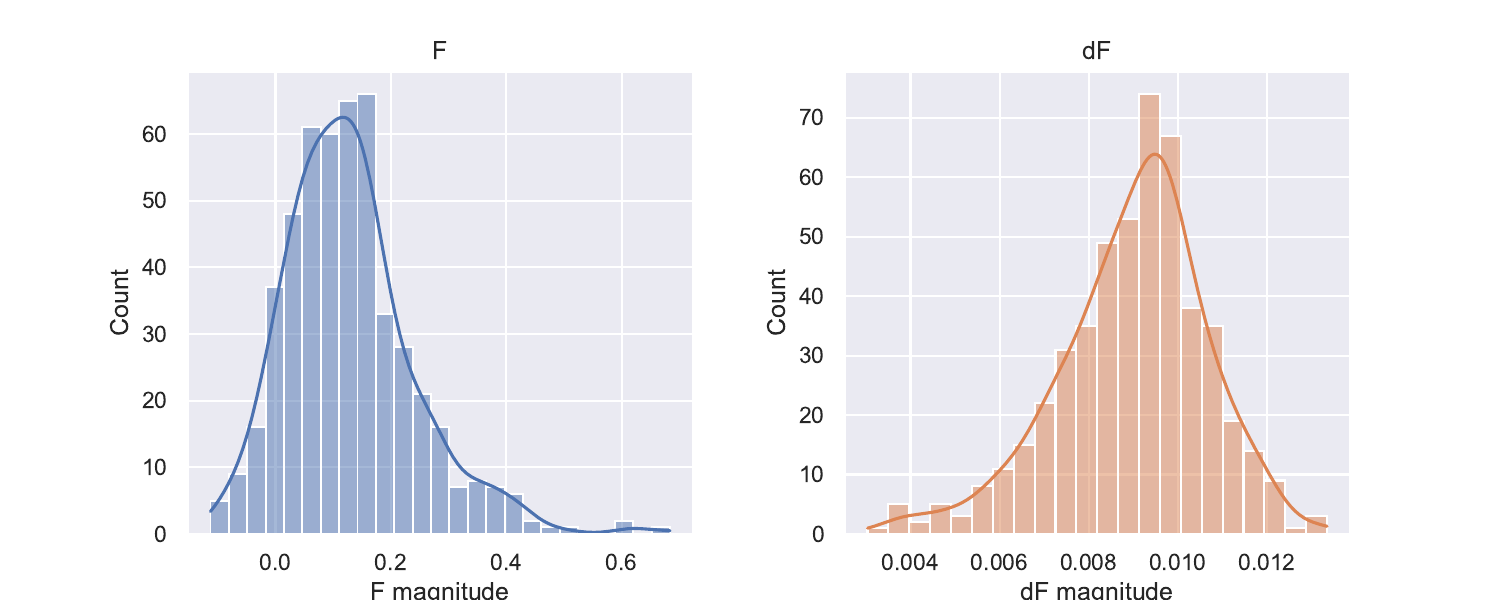}}
    \subfigure[Vapor fraction = 0.7 Numerical]{
        \label{figurecompare.sub.2}
        \includegraphics[width=1\textwidth]{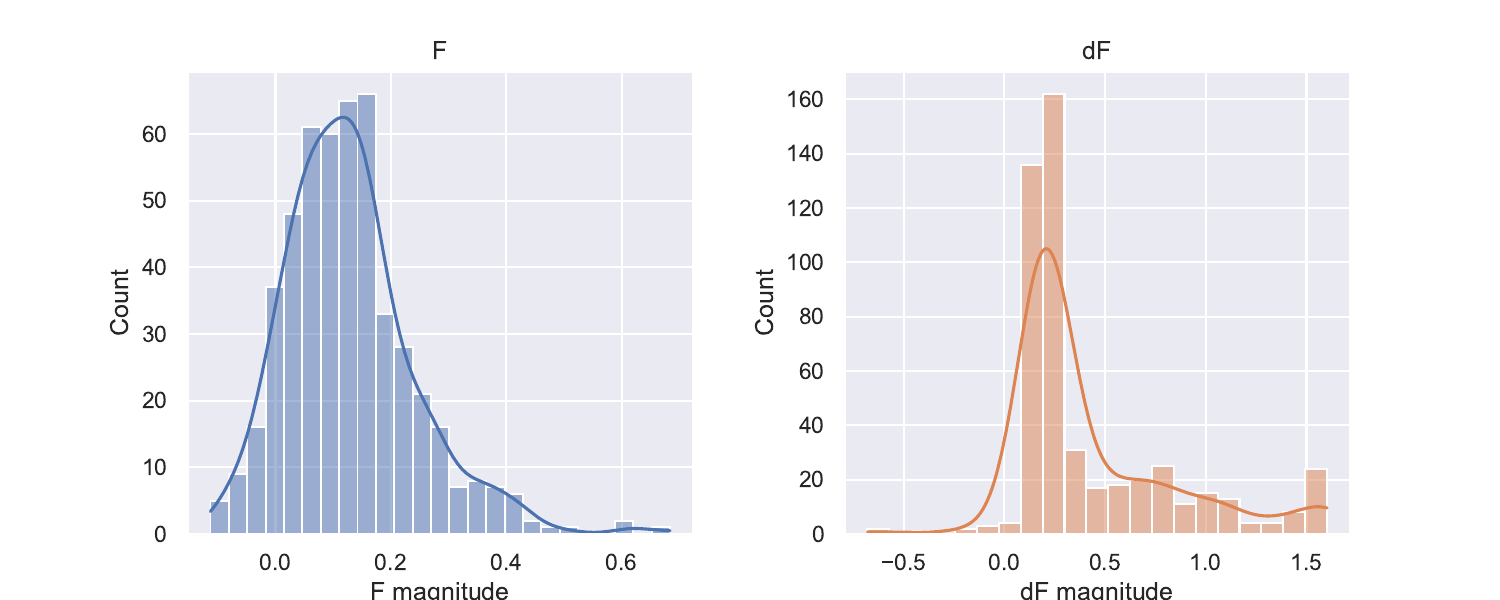}}
    \caption{Distribution of the convergence equation F derivatives generated by automated differentiation and numerical techniques at various temperatures}
    \label{figurecompare}
\end{figure}

The analysis of convergence is carried out mainly from the perspective of computational stability and the number of iterations of convergence.
In the stability analysis, 500 sets of experiments with randomly generated component fractions at a set vapor phase fraction of 0.7 are selected to analyze the distribution of $dFdT$ when PV flash is performed. Figure (\ref{figurecompare}) demonstrates that the distribution of derivatives obtained by auto-differentiation is more uniform, whereas the distribution of numerical derivatives obtained using finite differences has a greater variance and is more likely to result in computational failure when solving iteratively.\par
\begin{figure}[H] 
    \centering 
    \includegraphics[width=0.5\textwidth]{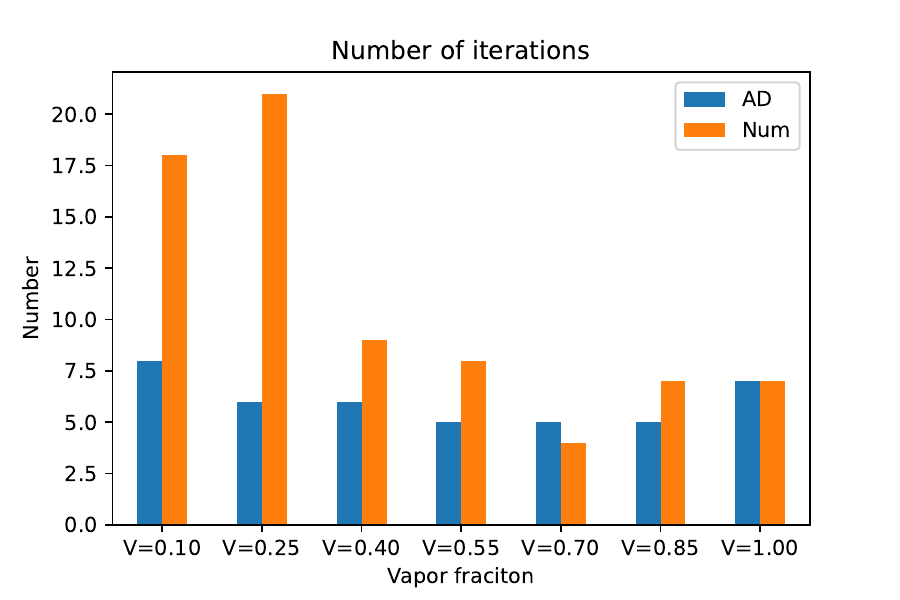} 
    \caption{Number of iterations} 
    \label{figure_iteration} 
\end{figure}

At the level of the number of iterations, we set different gas phase fractions to analyze the number of iterations of the flash calculation. Figure(\ref{figure_iteration}) illustrates the amount of iterations required to converge under various situations; the differentiable thermodynamic model requires fewer iterations, while the numerical differentiation approach requires more iterations. Because to the implicit formulas for the convergence equation, the AD technique provides clear numerical stability and convergence benefits over the numerical differentiation method. As long as the exact derivatives are producedKJthe calculation of the unit operation can converge better; but, if there are noticeable fluctuations in the derivatives, their convergence may lead to convergence into the error interval and the failure of the final calculation.\par
For the analysis of the simulation convergence failure, model calculation errors due to configuration errors were ruled out, and the main convergence problem of the model was traced to the fact that the thermodynamic phase equilibrium calculation iterations did not converge to satisfy the conditions of material conservation in each phase and system energy conservation. The most common phase equilibrium calculation in basic chemical unit operations is the temperature enthalpy (Duty) flash, whose core consists of adjusting the temperature, updating the thermodynamic properties of each phase based on the calculated temperature and phase equilibrium constant, and determining whether convergence conditions are satisfied.
% Figure () shows the pressure fraction, the derivative of the convergence target relative to temperature in pressure-loaded flash. 
% When using the auto-differential thermodynamic model, the exact derivative of the temperature with respect to the convergence target can be accurately reflected, and this derivative can accurately whether the process can converge or not. In the actual calculation, an adjustment mechanism can be introduced to fine-tune the input temperature conditions and recalculate the derivative when the calculation reveals significant fluctuations in the derivative to achieve self-correction of the model calculation and improve the stability of the result calculation.
% Using the auto-differentiation thermodynamic model, the precise derivative of the temperature with respect to the convergence objective may be exactly reflected, and this derivative can precisely determine whether or not a process can converge.

Using the auto-differentiation thermodynamic model, the precise temperature derivative with regard to the convergence goal may be properly reflected, and this derivative can precisely identify whether or not a process can converge.

% In the actual calculation, an adjustment mechanism can be introduced to fine-tune the input temperature conditions and recalculate the derivative when the calculation reveals significant fluctuations in the derivative, thereby achieving self-correction of the model calculation and enhancing the calculation's stability.
In the actual calculation, an adjustment mechanism can be implemented to fine-tune the input temperature conditions and recalculate the derivative when the calculation reveals significant fluctuations in the derivative, thereby achieving self-correction of the model calculation and improving the stability of the calculation.

\section{Conclusion}\label{section4}
In this study, we employ a cutting-edge framework for automatic differentiation within thermodynamic calculations, enabling the acquisition of precise derivatives without altering the algorithm's underlying logic. This approach stands in contrast to traditional numerical difference algorithms and results in substantial improvements in the convergence and computational efficiency of process simulation calculations. The benefits of the automatic differentiation method are assessed in the context of standard chemical phase equilibrium calculations, such as PT, PV, and PH flash, with a focus on numerical stability and iteration count.

Our experimental outcomes reveal that the automatic differentiation technique exhibits a more uniform gradient distribution and necessitates a reduced number of convergence iterations. Future research will extend the application of automatic differentiation in process simulation systems through three primary avenues: (1) augmenting the comprehensiveness of the tool library by developing additional thermodynamic method modules and unit operation computational models; (2) more effectively integrating the approach within the solution of large-scale process optimization design problems; and (3) enhancing the code's runtime efficiency by leveraging computational graph optimization techniques\cite{OptimizingDNN}.

\bibliography{mybibfile}

\end{document}